\documentclass[english,12pt]{article}

\ifx\pdftexversion\undefined\else
  \pdfoutput=1
  \usepackage[T1]{fontenc}
  \usepackage[utf8]{inputenc}
\fi

\usepackage{microtype}
\usepackage{babel}
\usepackage{lmodern}
\usepackage{csquotes}
\usepackage{fullpage}
\usepackage{mathtools}
\usepackage{amssymb}
\usepackage{hyperref}
\usepackage
 [style=numeric, sorting=none, giveninits=true, maxbibnames=10, backref=true]
 {biblatex}
\usepackage{cleveref}
\usepackage{tabu}
\usepackage{tikz-cd}
 \usetikzlibrary{matrix}

\numberwithin{equation}{section}
\newcommand\be{\begin{equation}}
\newcommand\ee{\end{equation}}

\newcommand\iu{\mathrm i} 
\newcommand\CalO{\mathcal O}
\newcommand\CalN{\mathcal N}
\newcommand\ii{\mathrm i}
\newcommand\BP{{\mathbb P}}
\newcommand\BR{{\mathbb R}}
\newcommand\BC{{\mathbb C}}

\newcommand\BZ{{\mathbb Z}}
\newcommand\CY{X}
\newcommand\TX{{{\mathcal T}_{\CY}}}
\newcommand\TNk{\ensuremath{TN_n}}

\begin{document}

\begin{center}
{\Large \bf Playing with the index of M-theory}\\
[12mm]
{\bf Michele Del Zotto${}^{a,b}$, Nikita Nekrasov${}^c$, \\
Nicol\`o Piazzalunga${}^b$ and Maxim Zabzine${}^b$}\\
[8mm]
{\small \it
   ${}^a$Mathematics Institute, Uppsala University,\\ Box 480, SE-75106 Uppsala, Sweden \\
   \vspace{.5cm}
   ${}^b$Department of Physics and Astronomy, Uppsala University,\\ Box 516, SE-75120 Uppsala, Sweden\\
   \vspace{.5cm}
   ${}^c$Simons Center for Geometry and Physics, Stony Brook University, \\
   Stony Brook NY 11794-3636, USA\\
   Center for Advanced Studies, Skoltech, Moscow, Russia\\
   Kharkevich Institute for Information Transmission Problems, Moscow, Russia\\
}
\end{center}
\vspace{7mm}

\begin{abstract}
Motivated by M-theory, we study rank $n$ K-theoretic Donaldson-Thomas theory on a toric threefold $X$.
In the presence of compact four-cycles, we discuss how to include the contribution of D4-branes wrapping them.
Combining this with a simple assumption on the (in)dependence on Coulomb moduli in the 7d theory,
we show that the partition function factorizes and,
when $X$ is Calabi-Yau and it admits an ADE ruling,
it reproduces the 5d master formula for the geometrically engineered theory on $A_{n-1}$ ALE space,
thus extending the usual geometric engineering dictionary to $n>1$.
We finally speculate about implications for instanton counting on Taub-NUT.
\end{abstract}

\eject
\tableofcontents


\section{Introduction}
\label{s:intro}

One can view the development of topological string theory as a journey
from world sheet to target space:
based on the realization \cite{Antoniadis:1993ze} that
the topological string free energy computes coefficients of effective action terms
in the graviphoton background,
the curve counting was re-interpreted \cite{Gopakumar:1998ii, Lawrence:1997jr, Gopakumar:1998jq}
in terms of BPS state counting in string/M-theory, coming from M2-branes,
with its genus-zero part giving a relativistic generalization
of Seiberg-Witten theory \cite{Nekrasov:1996cz}.
Later on a tool was developed to compute the topological string partition function/instanton partition function
in terms of box counting \cite{Nekrasov:2002qd, Nekrasov:2003rj, Aganagic:2003db, Iqbal:2007ii},
which led to the connection with Donaldson-Thomas theory \cite{Maulik:2004txy},
geometric engineering \cite{Katz:1996fh}, and spinning black holes \cite{Katz:1999xq}.

Usual DT theory is obtained by placing a single D6-brane on a threefold $X$ in type IIA string theory,
which in M-theory becomes the Taub-NUT space.
Similarly, for higher rank DT theory, we consider the $U(n)$ theory on the worldvolume of $n$ D6-branes wrapping $X\times S^1$.
In the limit where we send the Taub-NUT radius to infinity, we obtain the $A_{n-1}$-type ALE space.
At the same time a certain harmonic two-form that is $L^2$ on the Taub-NUT space becomes non-normalizable on the ALE space.
Correspondingly the associated $U(1)$ factor in the gauge group decouples.
If $\CY$ is a canonical Calabi-Yau three-fold singularity,
geometric engineering in M-theory assigns to it a five-dimensional superconformal field theory $\TX$.
Schematically,
\be
\begin{tikzcd}
Z^{7d}_{U(n)} (\CY \times S^1) \arrow[d]
\arrow[leftrightarrow, r,"\text{geom eng}"]
& Z^{5d}_{\TX} (TN_n \times S^1) \arrow[d,"R \to \infty"] \\
Z^{7d}_{SU(n)} (\CY \times S^1)
\arrow[leftrightarrow, r]
& Z^{5d}_{\TX} (A_{n-1} \times S^1)
\end{tikzcd}
\ee

Since $\TNk$ is non-compact, we can give boundary conditions at infinity to the scalar fields in $\TX$.
In particular, we can give vev to the operators parametrizing the Coulomb branch of $\TX$.
The latter correspond to the volumes of 2-cycles that arise from intersecting compact divisors in a smooth crepant resolution of $\CY$.
If $X$ is  non compact, we also have compact 2-cycles that arise from intersecting compact divisors with non-compact ones:
these correspond to mass deformations of $\TX$, which are the only susy preserving relevant deformations in 5d.
This is how the dependence on the K\"ahler parameters of $X$ enters the 5d partition function of $\TX$.
We summarize our notations/dictionary, which will be explained later.
\be
\begin{tabu}{l|l}
TN_n \times S^1_\beta & X \times S^1_\beta \\
\hline
5d & 7d \\
\TX & U(n) \\
q_a = e^{\beta \epsilon_a}, \quad
q_4,q_5 & q_1,q_2,q_3;p\\
\operatorname{rk} \TX (\text{Coulomb } b_\alpha=e^{\beta \varphi_\alpha}) &
\dim H_4(X,\BZ) \\
\operatorname{rk} \TX  + \operatorname{def} \TX (\text{inst }z, \text{masses}) &
Q_\alpha = e^{t_\alpha}, \alpha=1,\ldots,\dim H_2(X,\BZ) \\
\text{2-cycles} & \text{Coulomb } a_i=e^{\beta \alpha_i}, i=1,\ldots, n
\end{tabu}
\ee

The two main achievements of this paper are as follows:
\begin{itemize}
\item given any toric threefold $X$,
we extend usual Donaldson-Thomas theory in two directions:
first by going to higher rank, namely from $U(1)$ to $U(n)$ gauge theory;
second by including the contribution of D4-branes wrapping compact divisors.
A simple assumption on the dependence on equivariant parameters allows us to
prove a factorization property for this theory, which we call 7d master formula.
\item if X is also Calabi-Yau and admits a geometric engineering limit,
our 7d master formula matches the master formula
for the geometrically engineered 5d gauge theory on $A_{n-1}$ space,\footnote
{We work in K-equivariant setting, so the meaning of 7d is $S^1 \times X$.
A similar remark applies to 5d.}
which is the K-theoretic extension of usual 4d master formula.
\end{itemize}

Our motivation comes from M-theory (hence the title):
although we will not be able to provide a full derivation of everything from M-theory,
our construction has a clear 11d origin,
which suggests the equality between two protected quantities
as they come from different reductions of the same 11d object.
Conversely, our computations can be regarded as an equivariant test of M-theory.
Nevertheless, the main statements and conjectures of our paper can be formulated
in a mathematically rigorous way, ignoring their physical origin.

Our story is in many ways an extension of the work \cite{Nekrasov:2014nea},
where higher rank DT theory was presented,
and its connection to the index of M-theory on Calabi-Yau fivefolds was discussed.
We explore the effect of additional topological sectors,
allowing for sheaves with nontrivial $c_1$ on the threefold side,
and the fluxes through the $2$-cycles on the two-fold side.
Certain bits of our story appeared previously in the work \cite{ Nakajima:2015txa},
where the relation between the instantons on ALE and ALF spaces was studied,
and hints at a DT-like interpretation were pointed out.
Physically, our approach includes in a crucial way the effects of the $D4$-branes,
which were not considered in the abovementioned papers.

\subsection{Plan}

In \cref{s:M-theory} we review the M-theory background that underlies our computations.
Although some aspects of the story are well-known,
the full lift of the equivariant $\Omega$-background, including the $G_4$ flux,
that would allow to perform the localization calculations directly in M-theory,
is not.
Some of our considerations therefore remain qualitative.\footnote
{Some progress can be made along the lines of ref.~\cite{Costello:2016nkh}.}

In \cref{sec:5d} we review the instanton counting in 4+1d on non-compact toric manifolds,
in particular we present a straightforward extension to 4+1d of the 4d master formula.
We discuss the simplest cases, namely the vanishing Chern-Simons level and no matter,
but we believe our findings are valid more generally.
We also compare the ALE and ALF cases, and present a toy model computation in detail.

In \cref{sec:dtrev} we review the Donaldson-Thomas theory on a toric threefold $X$,
and extend it to the higher rank.
We recall useful facts from toric geometry and the DT/PT correspondence for local $X$.

In \cref{sec:dtnew} we combine the previous ingredients with the Coulomb-independence hypothesis,
and explain how to introduce the $D4$-branes.
The main result there is the 7d master formula.
This can be seen either mathematically as a factorization property for a generic toric threefold $X$,
or as an extension of the usual geometric engineering if $X$ is Calabi-Yau and engineers a gauge theory.
In the latter case, the 7d master formula matches exactly the 5d one for the corresponding theory.

While in \cref{sec:dtnew} we keep the discussion general,
in \cref{s:examples} we try to give as many details as possible for a few relevant examples.
After spelling out some details of the geometric engineering dictionary,
we test our findings on some of the geometries engineering
the $SU(N)$ gauge theory with zero CS level for $N=2,3$.

\section{M-theory setup}
\label{s:M-theory}

We review the M-theory framework that motivates our paper \cite{Dijkgraaf:2007sw,Nekrasov:2014nea}.
We begin with an overview of the general structure,
and then discuss the special class of backgrounds that give rise
to the examples we consider in this paper.

\subsection{An identity from Calabi-Yau fivefolds}

M-theory admits supersymmetric compactifications on Calabi-Yau 5-folds (CY5) of the form
\be
M_{11} = \BR \times M_{10}
\ee
which for generic CY5 preserve two supercharges \cite{Haupt:2008nu}.

In our paper we consider manifolds $M_{10}$ admitting isometries.
In this context we can define the twisted Witten index
\be
{\rm Tr} (-1)^{F} g = Z (S^{1} {\tilde\times} M_{10})
\label{eq:M-witten}
\ee
where $S^{1} {\tilde\times} M_{10}$ denotes a fiber bundle over $S^1$ with fiber $M_{10}$,
which is the cylinder of the isometry map $g: M_{10} \to M_{10}$.
We assume $g$ to commute with some supercharge.

Of course, for compact $M_{10}$, this makes no sense, since,
firstly, one is supposed to integrate over all metrics on $M_{10}$,
and secondly, all diffeomorphisms of $M_{10}$,
including the rare instances of isometries of a fluctuating metric,
are gauge symmetries, and, therefore, act trivially on the physical states.
Hence, we assume $M_{10}$ to be a non-compact space, asymptotically approaching
a fixed CY5 with nontrivial isometries.
These isometries are then treated as global symmetries.

We denote by $\mathcal T_{M_{d}}$ the $(11 - d)$-dimensional theory
obtained from M-theory on $\BR^{1,10-d} \times M_d$.
If $M_d$ is non-compact $\mathcal T_{M_{d}}$ is non-gravitational.
More precisely, the gravitational physics is fully eleven-dimensional,
while the dynamics of the $11-d$-dimensional (localized) degrees of freedom
takes place in the fixed gravitational background.
Actually, as explained in \cite{Iqbal:2003ds}
certain gauge-like degrees of freedom can be interpreted as topology changes,
thus representing the gravitational dynamics using supersymmetric gauge theory
(this could be compared to the AdS/CFT duality, in a topological context).

When $m+k=5$, the index \cref{eq:M-witten} can be interpreted in two ways:
on the one hand we have the partition function of $\mathcal T_{M_{2k}}$ on $S^1 \times M_{2m}$,
on the other we have the partition function of $\mathcal T_{M_{2m}}$ on $S^1 \times M_{2k}$.
These have to agree, giving the identity
\be
 Z^{(11-2k)d}_{\mathcal T_{M_{2k}}} (S^1 \times M_{2m})
 =Z ({S^1 \times M_{2k} \times M_{2m}})
 = Z^{(11-2m)d}_{\mathcal T_{M_{2m}}}(S^1 \times M_{2k})
\ee

\subsection{A 7d/5d correspondence}

The CY5 of our interest are a product
\be
M_{10} = M_4 \times M_6
\ee
where $M_4$ is either the charge $n$ Taub-NUT space
or an ALE space and $M_6=X$ is a CY3 singularity.\footnote
{Resolving the singularity gives rise to a flow to the Coulomb phase of the SCFT,
	which we denote $\CY_I$.
The index $I$ denotes possibly inequivalent resolutions of the singularity $M_6$
that correspond to different chambers in the Coulomb branch of the SCFT.
The corresponding geometries are birational smooth CY3 related by flop transitions.
Whenever $\CY_I$ admits a ruling supporting resolutions of ADE singularities,
that phase of the CB geometry can be interpreted in terms of gauge theory.
This is the case for the examples we consider in this paper,
and for this reason we often omit the subscript $I$ from $\CY_I$,
as we are considering an explicit gauge theory phase as our $\CY$.}
The $M_4$ spaces at their most singular point in the K\"ahler moduli engineer
7d maximally supersymmetric Yang-Mills theories in M-theory \cite{Sen:1997js}.
The space $\CY$ engineers a 5d SCFT $\TX$ in M-theory \cite{Morrison:1996xf, Intriligator:1997pq}.
The resulting geometries preserve 4 supercharges and
both give rise to non-gravitational theories.
We are led to an equation of the form
\be
Z^{7d}_{\mathcal T_{M_4}}(S^1_\beta \times M_6) =
Z^{5d}_{\mathcal T_{M_6}}(S^1_\beta \times M_4)
\ee
where the partition functions are interpreted as twisted Witten indices.
Since both spaces are non-compact,
these partition functions depend on choices of boundary conditions at infinity.

\subsection{A heuristic argument: topological bootstrap}

In the case $M_4 = \TNk$ we have a relation with higher rank DT theory,
building upon the classical duality among M-theory on $S^1_\beta \times \TNk \times X$
and IIA on $S^1_\beta \times \BR^3 \times X$ with $n$ D6 branes wrapping $S^1_\beta \times X$,
and exploiting the Taub-NUT circle as the M-theory circle.

One could add D4 branes wrapping $S^1_\beta \times D$,
where $D$ is a holomorphic 4-cycle of $X$,
\be
 \operatorname{vol} D = \int_{D} \omega \wedge \omega~,
\ee
where $\omega$ is the Kahler form of $X$.
These are non-supersymmetric at first sight:
indeed for $X = \BC^3$ one such state would correspond to a parallel system of D4-D6 branes,
which breaks supersymmetry as the number of Dirichlet-Neumann directions is not a multiple of 4.
However, in that context the D4-brane dissolves into flux for the D6-brane.
Therefore we could in principle include these configurations
at the price of dissolving the D4-branes into localized flux in our background.
Dualizing these D4-branes back to M-theory we obtain M5-branes wrapping the Taub-NUT circle,
which is fibered and shrinks at the position of the D6-branes.
These M5-branes are localized where the Taub-NUT circle shrinks
and dissolve in $G_4$ flux localized in the complement of such region.
Depending on how we do the reduction, we have two possible ansatzes
\be
\label{eq:D4part1}
\begin{aligned}
    G_4 &\sim F^{7d}_a \wedge  B^a  +  m_{i,a} PD_X[D^i] \wedge B^a \\
    G_4 &\sim F^{5d}_i \wedge PD_X[D^i] +  m_{i,a} PD_X[D^i] \wedge B^a
\end{aligned}
\ee
where $m_{i,a}$ is the number of M5 branes that are wrapped on $C_a$
(see \cref{App:TN} for notations) inside $\TNk$
and the other fields represent the KK modes corresponding to the field strengths
of the 7d and 5d theories, respectively.
Here $PD_X[D^i]$ stands for Poincar\'e dual of compact four cycles $D^i$ in $X$.
This suggests that M5-branes wrapping the Taub-NUT circle and a compact divisor within the CY3
can be interpreted as nontrivial first Chern classes for either of the curvatures
of the field theories in the 5d/7d correspondence.
Indeed, we have
\be
\label{eq:D4part2}
\begin{aligned}
    G_4 &\sim (F^{7d}_a + m_{i,a} PD_X[D^i]) \wedge B^a \\
    G_4 &\sim (F^{5d}_i +  m_{i,a} B^a) \wedge PD_X[D^i]
\end{aligned}
\ee
and each non zero $m_{i,a}$ can be absorbed as a non-trivial first Chern class
for the curvatures on the 7d and the 5d sides.
This discussion is purely heuristic and at the moment
we do not have enough tools to derive 5d/7d actions from 11d M-theory perspective.
However we know that the properly defined volume ${\cal F}(t)$ of CY
(triple intersection number of $X$) can be interpreted as
the prepotential of the rigid supersymmetric five-dimensional theory
\cite{Seiberg:1996bd, Morrison:1996xf, Intriligator:1997pq}.

The bootstrap approach to quantum field theory of \cite{Polyakov:1974gs}
recently has led to great advances in the quantitative analysis
of conformal field theories in three and four dimensions
(see e.g.~\cite{Rychkov:2020rcd, Poland:2018epd}).
The conformal bootstrap in two dimensions, at the level of a $4$-point correlation function
\be
\langle {\CalO}_{1}(x_1) {\CalO}_{2}(x_2) {\CalO}_{3}(x_3) {\CalO}_{4}(x_4) \rangle
\ee
is the requirement of the equality of two expansions,
one in the limit $x_{2} \to x_{1}$
(which is equivalent, thanks to conformal invariance, to $x_{3} \to x_{4}$ limit),
and another in the limit, e.g., $x_{3} \to x_{2}$
(equivalent to $x_{4} \to x_{1}$).
These expansions correspond to the respective $s$- and $t$-channel tree diagrams
(labelling the $4$-point conformal blocks).
In the context of toric geometry, similar tree diagrams describe
the two phases of the resolved conifold
$X = \left[ {\CalO}(-1) \oplus {\CalO}(-1) \longrightarrow {\mathbb{CP}}^{1} \right]$,
which can be described as the symplectic quotient of ${\BC}^{4}$ by $U(1)$:
\be
|z_{1}|^2 + |z_{2}|^2 - |z_{3}|^2 - |z_{4}|^2 = r \, , \qquad
\left( z_{1}, z_{2}, z_{3}, z_{4} \right) \sim
\left( e^{\ii\theta} z_{1}, e^{\ii\theta} z_{2}, e^{-\ii\theta}z_{3}, e^{-\ii\theta}z_{4} \right) \ .
\ee
For $r \neq 0$, the edges of the toric polytope $\Delta_{X}$
(not to be confused with the $1$-skeleton ${\Delta}_{X}^{(1)}$ used in this paper)
consist of four semi-infinite axes $l_{1}, l_{2}, l_{3}, l_{4}$ and one finite interval $c$.
For $r > 0$ these are
$l_{1} = \{ z_{3} = z_{1} = 0\} \cup l_{2} = \{ z_{4} = z_{1} = 0 \} \amalg l_{3} = \{ z_{3} = z_{2} = 0 \} \cup l_{4} = \{ z_{4} = z_{2} = 0 \}$,
and $c = \{ z_{3} = z_{4} = 0\}$, respectively.
For $r< 0$ the geometry is identical with $(z_{1}, z_{2}) \leftrightarrow (z_{3}, z_{4})$.
The generating function of Gromov-Witten invariants admits the analytic continuation $r \to - r$,
so that the essential part of the instanton counting agrees for $s$- and $t$-channels.
Perhaps closer in spirit to the bootstrap of CFT is the associativity WDVV equation
obeyed by the genus-zero Gromov-Witten invariants \cite{Kontsevich:1994qz}.

We call the conjectured equality of the $5d/7d$ perspectives the \emph{topological bootstrap}.
We imagine it also corresponds to some homotopy between
the ``large TN - small CY'' and the ``small TN - large CY'' geometries,
akin to the flop transition $r \ll 0 \to r\gg 0$ of the resolved conifold.
The validity of our conjecture strengthens the belief in the existence of the underlying $11d$ theory.

\section{5d theory on \texorpdfstring{$TN_n \times S^1_\beta$}{TNnXS1}}
\label{sec:5d}

We review and discuss the properties of 4d and 5d instanton partition functions
on non-compact manifolds with $T^2$-action.
In particular we are interested in non-compact toric ALE spaces of type $A_{n-1}$
and their cousins $TN_n$, the multi-Taub-NUT spaces.

Let us start with the basic setup.
In 4d a ${\cal N}=2$ gauge theory can be twisted and placed on arbitrary manifolds.
After twisting, the theory can be recast as a cohomological field theory,
which is known as Donaldson-Witten theory.
If the underlying manifold admits a $T^2$ action,
then one can define equivariant Donaldson-Witten theory.
Originally equivariant Donaldson-Witten theory has been discussed on ${\BC}^2$
\cite{Losev:1997tp, Moore:1997dj, Lossev:1997bz, Moore:1998et}
and this effort has resulted in the definition of the instanton partition function
\cite{Nekrasov:2002qd, Nekrasov:2003rj}.
For pure $U(N)$ ${\cal N}=2$ gauge theory on ${\BC}^2$,
the full partition function is given by
\be
 Z^{4d}_{U(N)} ({\BC}^2; z, \vec{\varphi}, \epsilon_4, \epsilon_5) =
 Z^{4d}_{\rm cl} Z^{4d}_{\rm 1-loop}
 \sum_{l=0}^{\infty} z^l ~{\rm vol}_l (\vec{\varphi}, \epsilon_4, \epsilon_5)~,
 \label{NP-4Dflat}
\ee
where $\operatorname{vol}_l (\vec{\varphi}, \epsilon_4, \epsilon_5)$ is
the equivariant volume of the moduli space of instantons of charge $l$ and
$Z^{4d}_{\rm cl}$, $Z^{4d}_{\rm 1-loop}$ stand for the classical and 1-loop parts correspondingly.
Here the parameters $(\vec{\varphi}, \epsilon_4, \epsilon_5)$
are the equivariant parameters for the $T^{N+2}$ action on the moduli space of instantons,
where $\vec{\varphi}$ stands for the constant gauge transformations
(one refers to them as Coulomb branch parameters)
and $(\epsilon_4, \epsilon_5)$ stand for $T^2$-rotations of ${\BC}^2$.
The parameter $z$ is an instanton counting parameter.
The 4d ${\cal N}=2$ gauge theory on ${\BC}^2$ has a natural 5d lift
to ${\BC}^2 \times S^1_{\beta}$
and the partition function corresponds to the index
\be
Z^{5d}_{U(N)} ({\BC}^2 \times S^1_\beta; z, \vec{b}, q_4, q_5) =
Z^{5d}_{\rm cl} Z^{5d}_{\rm 1-loop} \sum_{l=0}^{\infty} z^l ~{\rm ind}_l (\vec{b}, q_4, q_5)~,
\label{NP-5Dflat}
\ee
where ${\rm ind}_l (\vec{b}, q_4, q_5)$ stands for the equivariant index
of the Dirac operator on the moduli space of instantons of charge $l$
and $\vec{b} = e^{\beta \vec{\varphi}}$, $q_4 =e^{\beta \epsilon_4}$, $q_5= e^{\beta \epsilon_5}$.
The index ${\rm ind}_l$ can be written as an integral of
the equivariant A-roof genus over the moduli space of instantons.
In 5d one can add a Chern-Simons term.
The partition function on $\BC^2$ and $\BC^2 \times S^1_\beta$ has been generalized
to a wide class of $\CalN=2$ supersymmetric theories and
it has been studied extensively in different contexts,
see ref.~\cite{Tachikawa:2014dja} for a review.

The equivariant Donaldson-Witten theory can be defined on any four manifold $M_4$ that admits isometries
and the most interesting case is when $M_4$ admits a $T^2$ action.
There are two distinct cases of such theories: the case of non-compact and compact $M_4$.
Here we concentrate on the case of non-compact four manifold with $T^2$-action.
The 4d and 5d partition functions can be defined in the same way as in \cref{NP-4Dflat,NP-5Dflat}
if we know the explicit construction of the corresponding instanton moduli space.
On general grounds we expect the appropriate torus action on the instanton moduli space
(e.g., $T^{N+2}$ action for the $U(N)$ theory).
The main new feature is that the partition function may depend on more parameters
associated to extra labels related to the moduli spaces and the underlying geometry of $M_4$.
In the partition function different configurations are weighted by the classical term
\be
 \int_{M_4} e^{H + \omega}~ {\rm ch} (F) ~,
\ee
which in the path integral gets extended to the appropriate equivariant observable
(in 5d on $M_4 \times S^1_\beta$ we can also add Chern-Simons terms).
Here $\omega$ is an invariant symplectic form on $M_4$
and $H$ the corresponding Hamiltonian for the $T^2$-action.
In principle, one can construct more general observables
but this is not relevant for our discussion.

If $M_4$ is a toric variety then it can be glued from $\BC^2$ pieces.
The corresponding 4d master formula for non-compact toric varieties
\cite{Nekrasov:2003vi, Nakajima:2003pg, Gasparim:2008ri} takes the form
\be
 \boxed{
 Z^{4d}_{SU(N)} (M_4; z, \vec{\varphi}, \epsilon_4, \epsilon_5) =
 \sum_{(\vec{h}_1, \ldots, \vec{h}_p) \in \BZ^{(N-1)p}}
 \prod_{i=1}^k
 Z^{4d}_{SU(N)}
 \Big ({\BC}^2; z, \vec{\varphi} + \sum_{j=1}^k \phi^{(i)}_j \vec{h}_j , \epsilon^{(i)}_4, \epsilon^{(i)}_5 \Big )
 }
\label{4D-gluing}
\ee
where we are interested in $SU(N)$ gauge theory.
Here we deal with a smooth toric variety with $k$ fixed points under the $T^2$ action
and for every fixed point there exists a $T^2$-invariant open affine neighborhood
isomorphic to $\BC^2$,
with $\epsilon^{(i)}_4$, $\epsilon^{(i)}_5$ encoding the $T^2$-action at fixed point $i$.
The integers $\vec{h}_j$ ($j=1, \ldots, p= \dim H^2_c(M_4, \BZ)$) correspond
to the so-called fluxes,
which are labeled by compactly supported $H^2_c (M_4, \BZ)$ in every Cartan direction.
In \cref{4D-gluing}, the weights $\phi^{(i)}_j$ are constructed from toric data.
\Cref{4D-gluing} admits different refinements,
for example we can fix the holonomy at infinity,
in case a boundary of the toric space has non-trivial topology
(allowing different flat connections at infinity).
We aren't interested in such refinements and leave them aside.
Our main interest are $SU(N)$ gauge theories,
so we assume the traceless condition for $\vec{\varphi}$
and for every $\vec{h}_j$ with the appropriate invariant scalar product.

We follow the review \cite{Bruzzo:2014jza},
where one may find further mathematical details.
We assume that \cref{4D-gluing} has a straightforward 5d lift.
In 5d Chern-Simons terms can be introduced,
but we mainly ignore them to avoid cluttering in our formulas.

We are interested in two types of spaces:
ALE spaces of type $A_{n-1}$
and multi-Taub-NUT spaces $TN_n$,
which are both  hyperK\"ahler and admit $T^2$ isometries
(provided that the centres of these spaces are aligned).
Although $A_{n-1}$ is a limit of $TN_n$,
their instanton partition functions may differ,
since asymptotically they look different.
Let us start from the spaces $A_{n-1}$, which are examples of non-compact toric varieties.

\subsection{ALE spaces of \texorpdfstring{$A_{n-1}$}{An-1} type}

ALE spaces of type $A_{n-1}$ are hyperK\"ahler four-manifolds
that can be thought of as deformation (resolution) of the quotient $\BC^2/\BZ_n$,
with $\BZ_n$ being understood as subgroup of $SU(2)$ acting isometrically on $\BC^2$.
We collect some basic properties of $A_{n-1}$ spaces in \cref{App:An-1}.
In what follows we assume that the metric on $A_{n-1}$ has a $T^2$ isometry
and thus the centres are aligned.

There are two approaches to instanton partition functions on $A_{n-1}$.
In the first approach one constructs the instanton moduli space directly,
and this was done by Kronheimer and Nakajima \cite{MR1075769}
by considering ADHM data invariant under $\BZ_{n}$.
Later Nakajima \cite{Nakajima:1994nid} described them
in terms of Nakajima quiver varieties.
Thus one can define the instanton partition function on the $A_{n-1}$ space
as the  partition function for an appropriate quiver variety.
The second approach is based on the fact that the resolved $A_{n-1}$ space
is a toric variety and thus the full partition function on $A_{n-1}$
can be glued from $\BC^2$ pieces.
Physically the two approaches should produce the same result
as long as the partition function is independent from the sizes of resolved cycles.
However, this relation has not been proved, as far as we know.

Here we follow the second approach and assume that \cref{4D-gluing} gives
the full result for the $A_{n-1}$ space.
Our goal is to write the 5d version of this formula
with all toric data spelled out for $A_{n-1}$ (for a review see \cref{App:An-1}).
Gluing $A_{n-1}$ from $\BC^2$ pieces, the full 5d partition function takes the form
\be
\label{5d-master}
 Z_{SU(N)}^{5d} (A_{n-1} \times S^1; z, \vec{b}, q_4, q_5) =
 \sum_{(\vec{h}_1, \ldots, \vec{h}_{n-1}) \in (\BZ^{(N-1)})^{n-1}}
 \prod_{i=1}^n Z^{5d}_{SU(N)} ({\BC}^2 \times S^1; z, \vec{b}^{(i)}, q^{(i)}_4, q^{(i)}_5)
\ee
where we are ignoring the Chern-Simons level.
Here $q_4= e^{\beta \epsilon_4}$, $q_5= e^{\beta \epsilon_5}$ are
global parameters associated to the $T^2$ action,
while the local toric parameters
$q_{4}^{(i)}= e^{\beta \epsilon_4^{(i)}}$, $q_5^{(i)}= e^{\beta \epsilon_5^{(i)}}$
for fixed point $i$ are defined as
\be
 q_{4}^{(i)} = q_4^{n-i+1} q_5^{1-i}~,\quad q_5^{(i)} = q_4^{i-n} q_5^{i}~,
 \label{5ddef-q4-q5}
\ee
and these expressions can be read off from the toric data, see \cref{An-local-global}.
From global $\vec{b} = ({b_\alpha})$
with $\alpha = 1, \ldots, N$ being Cartan direction and $b_\alpha = e^{\beta \varphi_\alpha}$,
the local data are defined as
\be
 b_\alpha^{(i)} = b_\alpha (q_4^{(i)})^{h_{i,\alpha}} (q_5^{(i)})^{h_{(i-1),\alpha }} =
 b_\alpha (q_4^{n-i} q_5^{-i})^{h_{i,\alpha} -h_{(i-1),\alpha}} (q_4 q_5)^{h_{i,\alpha}}~,
 \label{def-b5d}
\ee
where $\vec{h}_i = \{h_{i, \alpha}\}$ are integers parametrized by
Cartan direction $\alpha$ and fixed point $i$.
Within geometric engineering, we are interested in $SU(N)$ theories,
thus in the above formulas we impose the trace condition
both for the Cartan parameters and for the fluxes, $h_{0,\alpha} =0 =h_{n,\alpha}$.


For the sake of our forthcoming discussion,
classical terms for $A_{n-1}$ geometry are glued as
\begin{multline}
 \beta^{-1} \log \Big (Z^{5d}_{\rm cl} (A_{n-1}\times S^1)\Big ) =
 \sum_{i=1}^n \frac
 {\Big \langle \vec\varphi + \vec{h}_i \epsilon_4^{(i)} + \vec{h}_{i-1} \epsilon_5^{(i)},
  \vec\varphi + \vec{h}_i \epsilon_4^{(i)} + \vec{h}_{i-1} \epsilon_5^{(i)} \Big \rangle }
 { \epsilon_4^{(i)} \epsilon_5^{(i)} } \\
 = \frac
 {\langle \vec\varphi, \vec\varphi \rangle }
 {n \epsilon_4 \epsilon_5}
 + \sum_{i=1}^n \Big ( 2 \langle \vec{h}_{i}, \vec{h}_{i-1}\rangle
 - 2 \langle \vec{h}_{i}, \vec{h}_i \rangle \Big )
 = \frac{\langle \vec\varphi, \vec\varphi \rangle}
 {n \epsilon_4 \epsilon_5} +{\cal C}_{ij} \langle \vec{h}_i ,\vec{h}_j\rangle
\label{5d-classical-An}
\end{multline}
where $\langle~,~\rangle$ stands for the Lie algebra pairing and
${\cal C}_{ij}$ is defined in \cref{matrix-An-1}
(it is related to the geometry of $A_{n-1}$).

\subsection{Multi Taub-NUT spaces \texorpdfstring{$TN_n$}{TNn}}

The cousins of ALE spaces of $A_{n-1}$ type are ALF spaces, the multi center Taub-NUT spaces $TN_n$.
They are four-dimensional hyperK\"ahler spaces asymptotic at infinity
to $\BR^3 \times S^1$, with $R$ the radius of this circle.
Close to the origin $TN_n$ looks like the $A_{n-1}$ space.
Thus $TN_n$ can be thought of as hyperK\"ahler deformation of $A_{n-1}$
with deformation parameter $R^{-1}$.
Taking $R$ to infinity reduces
the $TN_n$ hyperK\"ahler metric to the $A_{n-1}$ hyperK\"ahler metric.

As far as we are aware there is no formula for the instanton partition function on $TN_n$.
In 2008 Cherkis \cite{Cherkis:2008ip} initiated a systematic study
of the instanton moduli spaces for $U(N)$ gauge theory on $TN_n$.
The instanton moduli space on $TN_n$ is labeled by the following charges \cite{Cherkis:2010bn}:
the second Chern class $c_2$,
a collection of $n$ first Chern classes\footnote
{Or $(n-1)$ classes, there are some subtitles that we leave aside.}
$c_1$
and a collection of $N$ non-negative integer monopole charges $(j_1, \ldots, j_N)$.
The main novelty is the appearance of monopole charges related to the fact that
the self-duality condition is reduced to the monopole equation at infinity.
The bow diagrams (a generalization of quiver diagrams) encode
an ADHM-like construction for the moduli space of instantons \cite{Cherkis:2010bn}.
We are unaware of any direct equivariant calculation for this construction.
However, if we restrict to the zero-monopole sector
then the moduli space of instantons on $TN_n$ and on $A_{n-1}$ are related.
They are not isomorphic as hyperK\"ahler manifolds
but they are isomorphic as complex symplectic varieties \cite{Cherkis:2008ip, Witten:2009xu}.
Our guess is that,
since the partition function is not sensitive to the spacetime metric as long as the isometries are preserved,
the equivariant volume is the same for both spaces and thus
the instanton partition function for $TN_n$ in the zero monopole sector coincides
with the partition function for $A_{n-1}$.
In the next subsection we offer a toy calculation that may indicate this is true.
Again, the two spaces $TN_n$ and $A_{n-1}$ are different as hyperK\"ahler spaces,
but isomorphic as complex varieties, the isomorphism being $T^2$-equivariant.
We calculate the $T^2$-equivariant volume for both $TN_n$ and $A_{n-1}$
and show that they coincide.
This is an indication that a similar result is true for
the moduli spaces of $TN_n$ (zero monopole sector) and $A_{n-1}$.

\subsection{Toy calculation}

We evaluate the equivariant volume of $TN_n$ with respect to the $T^2$ action
and show that it agrees with that of $A_{n-1}$.
The original idea appeared in the work \cite{Moore:1997dj},
where part of the calculation was presented.
Here we spell out the details and use the full $T^2$ action on $TN_n$
with one $U(1)$ being the triholomorphic action
and another $U(1)$ the non-triholomorphic action
(for the metric to have these symmetries we require the centres to be aligned).

We follow ref.~\cite{Gibbons:1996nt} in the explicit construction of $TN_n$
as a hyperK\"ahler quotient.
With the standard quaternionic notations $i^2 = j^2 = k^2 = ijk = -1$,
let $\mathcal M = \mathbb H^n \times \mathbb H$,
with coordinates $q_a$ and $w$, for $a=1,\ldots, n$, with $G=\BR^n$ action
\be
q_a \to q_a e^{i t_a}~,
\quad
w \to w + R \sum_a t_a
\ee
with $R \in \BR$.
Take hyperK\"ahler quotient $TN_n = \mu^{-1}(\boldsymbol \zeta)/G$ with moment maps
\be
\mu_a = \frac12 \boldsymbol r_a + R \boldsymbol y~,
\ee
where $q_a = a_a e^{i \psi_a/2}$, $\boldsymbol r_a = q_a i \bar q_a$ and $w = y + \boldsymbol y$.
Here $y$ is real and $a_a,\boldsymbol y$ pure quaternions.
Let $\boldsymbol y = \frac{\boldsymbol r}{2R}$, $\boldsymbol \zeta_a = \frac12 \boldsymbol x_a$ and define
\be
\chi_a = \chi(\boldsymbol r_a) = \frac{da_a ia_a - a_ai da_a}{|a_a|^2}~,
\ee
so that $\chi = \sum_a \chi_a$ satisfies $d\chi = \star_3 dV$
with flat 3d metric and 
\be
V = \frac{1}{R^2} + \sum_{a=1}^n \frac1{|\boldsymbol x_a - \boldsymbol r|}~.
\ee
With $\tau = \sum_a \psi_a - \frac{2}{R} y$, the metric
\be
ds^2 = \sum_a dq_a \otimes d\bar q_a + dw \otimes d\bar w
\ee
becomes (after imposing moment map equations)
\be
ds^2 = \frac14 V d\boldsymbol r \otimes d \overline{\boldsymbol r} +
\frac14 \sum_{a=1}^n |\boldsymbol x_a - \boldsymbol r| (d\psi_a + \chi_a)^2 + dy^2~.
\ee
The vector fields generating the $G$-action are
\be
v_a = 2 \frac{\partial}{\partial \psi_a} + R \partial_y
\ee
and requiring the metric to satisfy $g(v_a,X)=0$ for any $a$ and $X$ yields
\be
|\boldsymbol x_a - \boldsymbol r| (d\psi_a + \chi_a) + 2 R dy = 0~.
\ee
Plugging this back, finally
\be
ds^2 = \frac14 V d\boldsymbol r \otimes d \overline{\boldsymbol r} + \frac14 V^{-1} (d\tau +\chi)^2
\ee
With $r_a = |\boldsymbol r_a|$, we have
\be
dq_a \wedge d \bar q_a = \frac{1}{4r_a} (r_a\chi_a-d\boldsymbol r_a)
\wedge (r_a\chi_a+d\boldsymbol r_a) +\frac12 d\psi_a \wedge d\boldsymbol r_a
\ee
so that Kahler forms
\be
 i \omega_I + j \omega_J + k \omega_K
= -\frac12 \sum_{a=1}^n dq_a \wedge d \bar q_a -\frac12  dw \wedge d \bar w
\ee
become (using moment maps)
\be
\sum_{a=1}^n dq_a \wedge d \bar q_a + dw \wedge d \bar w =
 -\frac14 V d\boldsymbol r \wedge d\boldsymbol r
 -\frac12 (d\tau +\chi) \wedge d\boldsymbol r
\ee
In complex coordinates $q_a = z_a + w_a j$, $\boldsymbol y  = x_r i + x_c k$ we have
\be
i \omega_I = i dy \wedge dx_r -\frac12 dx_c \wedge d\bar x_c
-\frac12 \sum_a dz_a \wedge d\bar z_a + dw_a \wedge d\bar w_a
\ee
while moment maps become
\be
\mu_a = i\left( \frac12 (|z_a|^2-|w_a|^2) + R x_r \right) + (R x_c-z_a w_a) k
\ee
The triholomorphic $U(1)_t$ acts as $\tau \to \tau + 2n \alpha$
with moment map $\mu_t = \frac{n}2 \boldsymbol r$.
If $\boldsymbol \zeta_a= i \zeta_a$ with $\zeta_a \in \BR$,
so that centers are aligned, there's
a non-triholomorphic $U(1)_n$ acting as
$q_a \to e^{i \alpha} q_a$, $w \to e^{i \alpha} w e^{-i \alpha}$,
which implies $z_a \to e^{i \alpha} z_a$, $w_a \to e^{i \alpha} w_a$,
$x_r \to x_r$, $x_c \to e^{2i\alpha} x_c$, with Hamiltonian
\be
H_n = |x_c|^2 + \frac12 \sum_a |z_a|^2+|w_a|^2
\ee
Up to a constant, the part of $\mu_t$ preserved by $U(1)_n$ is
\be
H_t = \sum_a \left( R x_r - \zeta_a \right)
\ee
and the equivariant volume is
\be
\operatorname{vol} (TN_{n}) :=
\int_{TN_n} dvol_g \exp (-\epsilon_n H_n -\epsilon_t H_t)
\ee
We have (using moment maps)
\be
r_a = |z_a|^2 + |w_a|^2
=  2 \sqrt{ (\zeta_a-R x_r)^2+| R x_c|^2 }
\ee
With $R x_c = \rho e^{i \theta}$,
we have
\be
\frac{\partial H_n}{\partial \rho} = 2 \rho V
\ee
and if we require $\Re \epsilon_n >0$ we see that the volume is independent of $R$ and it becomes
\be
\operatorname{vol} (TN_{n})=
\frac{2 \pi^2}{\epsilon_n} \int_{-\infty}^{+\infty} d\sigma
\exp (-\epsilon_n \sum_a |\sigma -\zeta_a| -\epsilon_t \sum_a (\sigma -\zeta_a))
\ee
Let's take $\Re \epsilon_n > | \Re \epsilon_t|$ and use analytic continuation.
By ordering $\zeta_1 < \zeta_2 < \ldots < \zeta_n$ we get
\begin{multline}
\operatorname{vol} (TN_{n}) / (4\pi^2)=
\frac1 {n(\epsilon_n - \epsilon_t)(\epsilon_n + \epsilon_t)}
- \frac12 \sum_{i=1}^n (\zeta_i-\frac{\zeta_*}n )^2 \\
- \frac{\epsilon_n}{3!} \sum_{i<j} (\zeta_i-\zeta_j)^3
+ \frac{n\epsilon_t}{3!} \sum_{i=1}^n (\zeta_i-\frac{\zeta_*}{n} )^3
+ O(\epsilon^2)~,
\label{vol-TN-exact}
\end{multline}
 where $\zeta_*= \sum_i \zeta_i$.
This agrees with \cref{volAn}, if we set
$\epsilon_t = \frac{1}{2} (\epsilon_5 - \epsilon_4)$,
$\epsilon_n = \frac{1}{2} (\epsilon_4 + \epsilon_5)$ and
$\zeta_a - \frac{\zeta_*}n = - \alpha_a$.
The first two terms agree with ref.~\cite{Moore:1997dj}.
The volume of $TN_n$ can be an inspiration for
the definition of 7d classical action \cref{our-def-classical}.

\section{DT theory on CY}
\label{sec:dtrev}

In this section, we review \cite{Okounkov:2015spn, Okounkov:2018yjl}
Donaldson-Thomas theory, focusing on toric Calabi-Yau\footnote
{
The Calabi-Yau condition is by no means necessary
from the viewpoint of DT theory on threefolds,
but it is useful when making contact with geometric engineering.
What needs to be CY is the five-fold.
}
threefolds $X$, and extend it to higher rank $n$. 
From a practical perspective, we view both
equivariant DT theory in 3 complex dimensions and equivariant Donaldson-Witten theory in 2 complex dimensions
as box counting problems \cite{Nekrasov:2014nea}.

\subsection{The setup}

Our type IIA setup consists of $n$ D6-branes (treated as background) wrapping $X \times S^1$,
with lower-dimensional branes wrapping cycles in $X$ and the circle,
in the presence of strong $B$-field along $X$.
The $(6+1)d$ non-commutative maximally supersymmetric $U(n)$ gauge theory \cite{Iqbal:2003ds}
on the D6 worldvolume leads at low-energy to quantum mechanics,
with target the instanton moduli space $\mathcal M$.
The K-theoretic DT partition function
\be
Z^{7d}_{U(n)}(X) = \sum_{ch} e^{u(ch)} \int_{[\mathcal M_{ch}]^{virt}}
e^{\omega+\mu_T} \hat A_T
\ee
is the generating function obtained by integrating $A$-roof genus on some virtual cycle.
We denote topological data $ch = ch (F)$ for some curvature $F$, and the classical factor
\be
\label{uclassqm}
u (ch) = \int_X e^{\omega + H} \sqrt{\hat A(X)} ch (F)
\ee
We denote $Z$ the summation restricted to $ch_1(F)=0$ and $\widehat Z$ the unrestricted one.
Integration is performed equivariantly with regard to a maximal torus $T$ of $U(3)\times U(n)$, parametrized by
$\Omega$-background parameters $q_1$, $q_2$, $q_3$ rotating  $X$
and Coulomb branch parameters $a_1,\ldots,a_n$ acting on the D6 Chan-Paton indices.\footnote
{We often suppress powers of $\beta$, the radius of $S^1$,
which can be restored by dimensional analysis.}
Each integral equals the twisted Witten index of the corresponding quantum mechanics.
The BPS objects contributing to the index are D0, D2 and D4 branes,
which wrap even-dimensional cycles in $X$ and can bound to D6-branes.
Localization reduces the computation to the fixed points of the action,
which are in correspondence with plane partitions.

\subsection{Toric data}
\label{toric-review}

We review basic facts and fix notations.
For $a=1,\ldots,N$ and $i=1,\ldots,n$, with $d = n-N>0$,
take a matrix $Q_a^i$ with integer entries,
and require that $\gcd (Q_a^1,\ldots,Q_a^n)=1$ for all $a$.
Let $t_a$ be positive real numbers.
On $\BC^n$ with coordinates $z_i$, define
momentum maps $\BC^n \to \BR^N$
\be
\mu_a(z)  = \sum_i Q_a^i |z_i|^2
\ee
Consider the set $\mu^{-1}(t) \subset \BC^n$ and take the quotient by $U(1)^N$ acting as
\be
z_i \to e^{\iu \sum_a Q_a^i \alpha_a} z_i
\ee
This is a subgroup of $U(1)^n$ acting as $z_i \to e^{\iu \varepsilon_i }z_i$.
The quotient is a $d$-dimensional toric variety $X$,
on which $U(1)^d = U(1)^n/U(1)^N$ acts with moment maps $\mu_H$,
which descend from
\be
H = \sum_i \varepsilon_i |z_i|^2
\ee
Similarly, the Kahler form $\omega$ on $X$ descends from the one on $\BC^n$,
and we have $\dim H_2(X) = N$.
Geometrically, choose a basis $C_a$ of $H_2(X)$.
The matrix $Q^i_a = D^i \cdot C_a$ represents intersection of
toric divisors $D^i = \{z_i=0\}\cap X$ with curves $C_a$,
and $t_a= \int _{C_a}\omega$.
We are interested in $d=3$ and $X$ Calabi-Yau, which implies $\sum_i Q_a^i = 0$.

To a toric threefold $X$, we can associate its polyhedron $\Delta_X$,
given by the image of $\mu_H$.
This has real dimension 3, and it is non-compact if $X$ is non-compact.
We call \emph{vertices} its zero-dimensional faces, $v\in \Delta_X^{(0)}$,
the fixed points of the $U(1)^3$ action discussed above.
Every vertex has valence 3,
namely there are 3 fixed lines (some of which can be non-compact) emanating from it.
Restricting to the compact skeleton of $\Delta_X$,
we call \emph{edges} the one-dimensional faces, $e \in \Delta_X^{(1)}$,
and \emph{faces} the two-dimensional ones, $f \in \Delta_X^{(2)}$.
Denote by $n_f$ the number of faces.
Generically, the number of edges in $\Delta_X^{(1)}$ is larger than $N$.

Around each vertex $v \in \Delta_X^{(0)}$, we can choose local coordinates,
made out of $U(1)^N$-invariant combinations of  $z_i$ variables.
These are acted upon by $U(1)^d$, their weights being the local $\Omega$-background parameters
(aka twisted masses in the GLSM language),
denoted by $q_1^{(v)}$, $q_2^{(v)}$, $q_3^{(v)}$ for $v \in \Delta_X^{(0)}$,
with $q_a^{(v)} = e^{\beta \epsilon_a^{(v)}}$.
They are functions of the global $\varepsilon$'s and transform in the same way
as the local coordinates, so only one such set is independent:
we denote it by $q_1$, $q_2$, $q_3$.
There's no canonical choice for such $q_1$, $q_2$, $q_3$.
The CY condition reads $q_{123}:=q_1 q_2 q_3 =1$, but we do not need to impose it.
We will often leave the label $(v)$ implicit and
denote $P_{123}=(1-q_1)(1-q_2)(1-q_3)$, $P_a=1-q_a$ for $a=1,2,3$.

For our gauge-theoretic purposes,
we associate an integer $m_f$ to each $f \in \Delta_X^{(2)}$
(this integers correspond to $c_1(F)$ of the 6d curvature $F$.)
From the viewpoint of a vertex, there are three such integers,
associated to the three faces this vertex sees
(with the understanding the $m=0$ for a non-compact face).
Let
\be
(q^{(v)})^m = (q_1^{(v)})^{m_{23}} (q_2^{(v)})^{m_{13}} (q_3^{(v)})^{m_{12}}
= e ^{ \beta \epsilon^{(v)} \cdot m}
\label{locmde}
\ee
where we identify direction 1 with face along 23, etc.
If $e \in \Delta_X^{(1)}$ connects vertices $v_1$ and $v_2$, then we have
\be
\epsilon_\tau^{(v_2)} = - \epsilon_\tau^{(v_1)},
\quad
\epsilon_{n_1}^{(v_2)} = \epsilon_{n_1}^{(v_1)} - \psi_{n_1}^{(e)} \epsilon_\tau^{(v_1)},
\quad
\epsilon_{n_2}^{(v_2)} = \epsilon_{n_2}^{(v_1)} - \psi_{n_2} ^{(e)}\epsilon_\tau^{(v_1)}
\ee
for some integers $\psi_{n_1}^{(e)}$ and $\psi_{n_2} ^{(e)}$.
(Here $\tau$ is for tangent, $n_1$ and $n_2$ for normal directions to the edge.)
In other words, $e \sim \BP^1$ and its normal bundle in $X$ splits as
\be
\CalN = \CalO (-\psi_{n_1}^{(e)}) \oplus \CalO (-\psi_{n_2}^{(e)})
\ee
If $X$ is CY, then $\psi_{n_1}^{(e)}+\psi_{n_2} ^{(e)}= -2$.
We define
\be
\label{edge-face}
\psi^{(e)} \cdot m =
\sum_{v \in e} \frac{\epsilon^{(v)} \cdot m}{\epsilon^{(v)}_\tau}
\ee
the sum being over the two vertices that belong to $e$.
This equals
\be
\psi^{(e)} \cdot m =
\psi^{(e)}_{n_1} m_{n_1} + \psi^{(e)}_{n_2} m_{n_2} + \sum_{v\in e} m_{\tau}
\ee
Again, the sum is over the two vertices that belong to $e$,
and $m_\tau$ refers to the face with normal direction $\tau$ at $v$.
This is cumbersome (but well-defined),
and we'll make it more geometric in a moment.
Given a Young diagram $\lambda$ (see below), we define
\be
f_\lambda^{(e)} = \sum_{(a,b) \in \lambda}
\psi_{n_1}^{(e)} \left( a-\frac12 \right) + \psi_{n_2}^{(e)} \left( b-\frac12 \right)
\ee
Denote by $t_e = \sum_{v \in e} \frac{H_v}{\epsilon_e^{(v)}}$ its size
and $Q_e = e^{t_e}$.

\subsubsection{From local to global}

The work \cite{wip} studies a map\footnote
{Recall that, for a three-fold $X$, $\dim H_4(X) = \dim H^2 (X,\BZ)_c$ by Poincaré duality,
where we view compact support cohomology as $H^2(X,\BZ )_c \subset H^2_{dR}(X)_c$.}
from $H^2_{c} (X)$ to $H^2 (X)$
\be
\label{gpsi}
m = (m^i)_{i \in \Delta_X^{(2)}} \mapsto
(\psi.m)^a := \sum_{i \in \Delta_X^{(2)}} Q_i^a m^i,
\quad a=1,\ldots,N
\ee
In that context, the geometry behind \cref{edge-face,locmde} is clear:
they are local versions of the global map just defined.
Borrowing certain definitions\footnote
{We temporarily switch to upper index $a$ and lower index $i$,
to match notations of that paper.}
and results from there,
we explain why this is the case.

Consider the K-equivariant integral\footnote
{The $t=0$ limit, which features e.g.~in \cref{calP}, gives
\be
Z_q (0) - Z_q (\psi.m) \prod_{i \in \Delta_X^{(2)}} q_i^{m^i}
= \sum_{v\in\Delta_X^{(0)}} \frac{1-q^{m}}{P^*_{123}}
\ee}
\be
\label{qdiff}
Z_q (t) - Z_q (t+\psi.m) \prod_{i \in \Delta_X^{(2)}} q_i^{m^i} =
\oint_{JK} d\phi \, e^{\phi.t}
\frac
 {1-\prod_{i \in \Delta_X^{(2)}} e^{-\beta x_i m^i}}
 {\prod_{i=1}^n (1-e^{-\beta x_i})}
\ee
The relation between Chern roots $x_i := \varepsilon_i + \sum_a Q_i^a \phi_a$
and local $\epsilon^{(v)}_{1,2,3}$ at a fixed point $v \in \Delta_X^{(0)}$
is such that, at any JK pole, all $x_i$'s are zero,
except for three (in this paper $d=3$),
from which we can read off the local $\epsilon^{(v)}$'s.
Moreover, each $\epsilon^{(v)}_{a=1,2,3}$ couples to
the face $f \in \Delta_X^{(2)}$ touching $v$ and with normal direction $a=1,2,3$,
precisely as in \cref{locmde}.
From this, it follows that \cref{edge-face} is induced by \cref{gpsi},
as implicitly assumed below.

All these properties are explicitly checked in the examples below.

\subsection{Partitions}

We can think of higher-dimensional partitions recursively. Start from a Young diagram:
this is a collection $\lambda=(\ell_1,\ldots,\ell_s)$ with $s \geq 1$ of positive integers $\ell_i$
such that $\ell_i \geq \ell_{i+1}$ for $i=1,\ldots,s-1$,
and we denote its size by $|\lambda|=\sum_{i=1}^s \ell_i$.
Inclusion is defined as $\lambda \subseteq \lambda'$ iff $\ell_i \leq \ell_i'$ for all $i$.
The next step is a plane partition: this is a collection $\pi=(\lambda_1,\ldots,\lambda_s)$
of Young diagrams $\lambda_i$ such that $\lambda_{i+1} \subseteq \lambda_i$.
Inclusion is defined as $\pi \subseteq \pi'$ iff $\lambda_i \subseteq \lambda_i'$ for all $i$'s,
and the size is $|\pi|=\sum_{k=1}^s |\lambda_k|$.
Equivalently we can think of a plane partition $\pi$ as a collection of
non-negative integers $\{ \pi_{i,j} \}$ indexed by integers $i,j \geq 1$ subject to the condition
\be
\pi_{i,j} \geq \max( \pi_{i+1,j}, \pi_{i,j+1}) \quad \forall i,j
\ee
The size is $|\pi| = \sum_{i,j} \pi_{i,j}$.
In this formulation, we can regard the plane partition $\pi$
as the subset of points $(a,b,c) \in \BZ^3$, such that $a,b,c \geq 1$ and $c \leq \pi_{a,b}$.
Its character is
\be
K_\pi (q_1, q_2, q_3)= \sum_{(a,b,c) \in \pi} q_1^{a-1} q_2^{b-1} q_3^{c-1}
\ee
A colored plane partition $\vec \pi = (\pi_1,\ldots,\pi_n)$ is
a $n$-dimensional vector of plane partitions, where we call $n$ the rank.
With $K_i = K_{\pi_i}$, we define its character as
\be
K = \sum_{i=1}^n a_i K_i (q_1, q_2, q_3)
\ee
Its size is $|\vec \pi|=\sum_{i=1}^n |\pi_i|$.
We define the dual $K^*$ of $K$ by replacing $q_a$ with $q_a^{-1}=q_a^*$ for $a \in \{1,2,3\}$
and similarly for $a_i$.
We will often identify a plane partition with its character.

\subsubsection{Regularization}

The partitions are allowed to have infinite size.
In this case, it is better to think of a partition $\pi$
in terms of the associated monomial ideal $I_\pi \subset \BC[q_1,q_2,q_3]$,
\be
\pi = \{ (k_1,k_2,k_3) \in \BZ^3_{>0} | \,\prod_{a=1}^3 q_a^{k_a-1} \not \in I_\pi \}
\ee
The asymptotics of $\pi$ along direction $a$ is given by
\be
\lambda_a = \lim_{q_a \to 1}  P_a \pi
\ee
and depends on all three variables except $q_a$.
The regularized partition is defined as
\be
K_{reg} = K - \sum_{\alpha=1}^3 \frac{\lambda_\alpha}{P_\alpha}
\ee

In analogy with partitions, we define the size of a Laurent polynomial $\mathcal P(q_1,q_2, q_3)$ as
\be
|\mathcal P| = \mathcal P(1,1, 1)
\ee
which can be negative.

\subsubsection{Plethystic substitutions}

A Laurent polynomial in the variables $q_a$ and $a_i$ is \emph{movable}
when it does not contain $\pm 1$ factors in the sum.
The map $\hat a$ is defined on movable Laurent polynomials as
\be
\hat a : \sum_i p_i M_i \mapsto
\prod_i \left( M_i^{1/2}-M_i^{-1/2}\right)^{-p_i}
\ee
where $M_i$ are monomials with unit coefficient and $p_i$ integers.

\subsection{Vertex formalism}

For generic $X$, fixed points are in one-to-one correspondence with collections $I$ of
$n$-tuples of (possibly infinite size) plane partitions, located at the vertices of $\Delta_X$:
\be
I = \{ \pi_v = (\pi_{1,v},\ldots, \pi_{n,v}) \}_{v \in \Delta_X^{(0)}}
\ee
Each $\pi_{i,v}$ is a plane partition,
and the collection satisfies certain compatibility conditions:
$\pi_{i,v_1}$ and $\pi_{i,v_2}$ must have the same asymptotics along edge $e$,
whenever $v_1$ and $v_2$ belong to $e$.

With $K_{i,v} = K_{\pi_{i,v}}(q_1^v,q_2^v,q_3^v)$ for  $v\in\Delta_X^{(0)}$,
the virtual tangent space at $I$ is
\be
T_I = \sum_{v\in\Delta_X^{(0)}} -P_{123} \mathcal H_v \mathcal H^*_v - T_{pert}
\ee
where we defined
\be
\mathcal H_v = \sum_i \frac{a_i}{P_{123}} - a_i K_{i,v}
\label{calH}
\ee
and subtracted the (divergent) perturbative factor
\be
T_{pert} = \sum_{v\in\Delta_X^{(0)}} -\frac{1}{P_{123}^*} \sum_{i,j} \frac{a_i}{a_j}
\ee
We can rewrite this as
\be
T_I =   \sum_{i,j} \frac{a_i}{a_j} N_{ij}
\ee
where we defined
\be
N_{ij} = \sum_{v\in\Delta_X^{(0)}} K^*_{j,v} -q_{123} K_{i,v} -P_{123}K_{i,v} K_{j,v}^*
\ee

Since the partitions can only grow along compact cycles, we know that $T_I$ is a Laurent polynomial,
and we are allowed to apply the $\hat a$ functor to it.
The partition function, aka twisted Witten index, takes the form
\be
Z^{7d}_{U(n)} (X) = \sum_I \hat a (T_I) \, e^{u(I)}
\ee
Let us redistribute \cite{Maulik:2003rzb, Maulik:2004txy} the various parts,
such that each one is manifestly finite.

\subsubsection{No faces}

Let us consider the case with no faces.
By using the regularized expression $K_{reg}$, we can write
\be
N_{ij} = \sum_{v\in\Delta_X^{(0)}} T_{v,ij} + \sum_{ v\in\Delta_X^{(0)}} \sum_\alpha t_{\alpha,ij}
\ee
where the first term contains regularized contributions and all other finite pieces
\begin{multline}
\label{tvtx}
T_{v,ij} =
K_{j,v,reg}^* - q_{123} K_{i,v,reg} -P_{123} K_{i,v,reg}K_{j,v,reg}^* \\
-P_{123} K_{i,v,reg} \sum_\alpha \frac{\lambda_{j,\alpha}^*}{P_\alpha^*}
-P_{123} K_{j,v,reg}^* \sum_\alpha \frac{\lambda_{i,\alpha}}{P_\alpha}
-P_{123} \sum_{\alpha\neq\beta} \frac{\lambda_{i,\alpha}}{P_\alpha}\frac{\lambda_{j,\beta}^*}{P_\beta^*}
\end{multline}
while the second term contains the infinite partitions,
\be
t_{e,ij} = 
\frac{\lambda_{j,e}^*}{P_e^*}
-q_{123} \frac{\lambda_{i,e}}{P_e}
-P_{123} \frac{\lambda_{i,e}}{P_e}\frac{\lambda_{j,e}^*}{P_e^*}
\ee
and it produces a finite term
\be
\label{tedge}
T_{e,ij} = \sum_{v \in e} t_{e,ij}
\ee
once we sum over the two vertices belonging to the edge.
Both $T_v$ and $T_e$ are movable Laurent polynomials.
(So we can apply plethystic to them.)
We have
\be
T_I = \sum_{i,j} \frac{a_i}{a_j} \sum_{v\in\Delta_X^{(0)}} T_{v,ij} + \sum_{e \in \Delta_X^{(1)}} T_{e,ij}
\ee

We apply Duistermaat-Heckman theorem to compute $ch=(ch_0,ch_1,ch_2,ch_3)$
\be
\label{classical}
\begin{aligned}
ch_3  &= \sum_i \left( \sum_{v \in \Delta_X^{(0)}}
 \left( -\frac{\alpha_i^3}{3! \epsilon_1^{(v)} \epsilon_2^{(v)}\epsilon_3^{(v)}} + |K_{i,v,reg}|\right)
	- \sum_{e \in \Delta_X^{(1)}} f_{\lambda_{e,i}}
 - \sum_{e \in \Delta_X^{(1)}} |\lambda_{i,e}| \sum_{v \in e} \frac{\alpha_i}{\epsilon_e^{(v)}} \right)
\\
ch_2 &= - \sum_i \left( \sum_{v\in\Delta_X^{(0)}}
\frac{\alpha_i^2 H_v}{2 \epsilon_1^{(v)} \epsilon_2^{(v)}\epsilon_3^{(v)}}
+ \sum_{e \in \Delta_X^{(1)}} |\lambda_{i,e}| t_e \right) \\
ch_1 &= -\sum_i \sum_{v\in\Delta_X^{(0)}} \frac{\alpha_i H^2_v}{2 \epsilon_1^{(v)} \epsilon_2^{(v)}\epsilon_3^{(v)}} \\
ch_0 &= - n \sum_{v \in\Delta_X^{(0)}} \frac{H^3_v}{3! \epsilon_1^{(v)} \epsilon_2^{(v)}\epsilon_3^{(v)}}
\end{aligned}
\ee
The last term in $ch_3$ is zero for the present case, but will contribute when we turn on fluxes.
The quantum mechanical expression \cref{uclassqm} is obtained by setting $\alpha=0$.
We get\footnote
{Compared to \cref{uclassqm}, we introduce higher times $\tau_p ch_p$,
which are discussed in \cref{s:class}, together with a proper treatment of $ch_0$.
From $\sqrt{\hat A(X)}$ or $\hat \Gamma$-class,
we only keep the term $-\frac{1}{24}c_2(X)\cdot t$.}
\be
u^{U(n)}_{K,\lambda} (g,t)
= g ch_3 + ch_2 - n \frac{1}{24} c_2(X) \cdot t + \frac{ch_0}{g^2 - (n\epsilon/2)^2}
\label{our-def-classical}
\ee
where we denoted
\be
\epsilon = \epsilon_1 + \epsilon_2+\epsilon_3 + 2\iu \pi
\label{ecy}
\ee

With $-p = e^{g}$, we split the sum over $I$ as a sum over $\pi$'s with given asymptotics $\lambda$ (vertex)
\be
V_{v,\lambda}=  \sum_{\pi|\lambda} (-p)^{\sum_i |K_{i,v,reg}|} \prod_{i,j} \hat a (\frac{a_i}{a_j} T_{v,ij})
\ee
and a sum over asymptotics, with the simple (edge) functions
\be
E_e (\lambda) =
(-p)^{-\sum_i f_{\lambda_{e,i}}} Q_e^{-\sum_i |\lambda_{i,e}|}
\prod_{i,j} \hat a (\frac{a_i}{a_j} T_{e,ij})
\ee
At this stage, there's no clear relation between the $n=1$ and $n>1$ cases,
which depend in a complicated way on Coulomb branch parameters. We get
\be
\label{7dunnom}
Z^{7d}_{U(n)} (X;p,t_e,a_i) =
e^{-\frac{n}{24} c_2(X) \cdot t + \frac{ch_0}{g^2-(n\epsilon/2)^2}}
\sum_\lambda \prod_{v\in\Delta_X^{(0)}} V_{v,\lambda} \prod_{e\in\Delta_X^{(1)}} E_e(\lambda_e)
\ee

\subsection{Rank one vertex and GV/PT}

Let $X$ be a non-compact toric threefold.
Up to a technical assumption, if we normalize the rank one vertex by the empty vertex,
the individual dependence on $q_1$, $q_2$, $q_3$ goes away \cite[Section 7.1.3]{Nekrasov:2014nea}
and the result only depends on their product.
The overall factor in \cref{7dunnom} is such that, for a geometry $X$ engineering theory $\TX$,
we get exactly \cite{Iqbal:2003zz, Eguchi:2003sj, Pandharipande:2007kc}
the full 5d instanton partition function of $\TX$ featuring in \cref{NP-5Dflat}:
\be
\label{swear}
\frac{Z_{U(1)}^{7d}(X)}{\prod_{v \in \Delta_X^{(0)}} Z_{U(1)}^{7d}(\BC^3) } = Z^{5d}_{\TX}(\BC^2)
\ee
provided $\Omega$-background parameters on the two sides are properly identified.


\section{General theory}
\label{sec:dtnew}

In this section we develop the general higher rank theory.
We first state our main assumption, and then work out its implications.
We first deal with the simpler case of no D4-branes,
and then add D4-branes wrapping the hypersurfaces of $X$, corresponding to the faces in $\Delta_X$.
For both cases, we derive a 7d master formula where the partition function completely factorizes.
For geometries admitting a geometric engineering limit,
this factorization reproduces exactly the 5d master formula on the corresponding $A_n$ space.
The focus here is on general results, while some examples are presented in the following section.

\subsection{Key assumption}

We \emph{assume} independence on Coulomb moduli in the instanton sector.\footnote
{Our assumption is actually a theorem for $X=\BC^3$ \cite{Awata:2009dd, Fasola:2020hqa}.}
Mathematically, this independence mirrors
the independence of equivariant parameters in \cite{Nekrasov:2014nea},
which is related to compactness of the corresponding moduli spaces,
but we take it as an experimental fact.
Again, all we need is the toric Calabi-Yau fivefold, so we can work, for example,
with $U(n)$ theory on $\BP^3$
(which is engineered \cite{Nekrasov:2014nea} by taking a resolution of singularities of
the total space of the direct sum of two ${\BZ}_{n}$-quotient of the sum of two line bundles,
i.e. ${\CalO}(-2) \oplus {\CalO}(-2)$).
We also don't need to be within the realm of the geometric engineering
in the sense of \cite{Katz:1996fh},
e.g.~we can analyze the theory on the total space of the line bundle $\CalO (-3) \to \BP^2$.

We performed several experimental checks of our assumption
both in the zero-flux sector, and when $c_1(F) \neq 0$ (highly non-trivial).

Physically, this independence is the independence of the partition function of the
$\Omega$-deformed five-dimensional ${\CalN}=1$ supersymmetric theory
on $\widetilde{{\BC}^{2}/{\BZ}_{n}}$ fibered over $S^1$,
on the K\"ahler moduli of the resolution.
This is the usual argument of the $Q$-exactness of
the appropriate components of the stress-energy tensor.
This means that the DT partition function can depend on seven-dimensional Coulomb moduli
only via an overall universal factor, which we suppress in the following.

\subsection{Factorizations}

Using notations and conventions of \cref{tvtx,tedge}, recall that
\be
N_{ij} = \sum_{v\in\Delta_X^{(0)}} T_{v,ij} + \sum_{e\in\Delta_X^{(1)}} T_{e,ij}
\ee
Observe that $N_{ji} = - q_{123} N^*_{ij}$.
With $N_i = N_{ii}$, write
\be
\sum_{i,j} \frac{a_i}{a_j} N_{ij} = \sum_i N_i + \sum_{i<j} \frac{a_i}{a_j} N_{ij} + \frac{a_j}{a_i} N_{ji}
\ee



Because of the assumption, we can take whatever choice of $a_i$,
and taking $a_i=L^i$ and then sending $L \to \infty$
is particularly convenient.


Set $a_i = L^i$ and look at the limit $L \to \infty$.
For any monomial $x$, we have
\be
\hat a (-L^{j-i} x)
=(L^{j-i}x)^{1/2} (1-L^{i-j}x^{-1})
\ee
With $i<j$, taking the conjugate of last term, we compute
\be
\lim_{L\to\infty} \hat a ( \frac{a_j}{a_i}  N_{ji} + \frac{a_j}{a_i}  N_{ij}^*)
= \prod_{v\in\Delta_X^{(0)}} q_{123}^{\frac12 |K_{j,v,reg}|-\frac12 |K_{i,v,reg}|}
\prod_{e\in \Delta_X^{(1)}} q_{123} ^{\frac12 (f_{\lambda_{i,e}} - f_{\lambda_{j,e}})}
\ee
where quadratic pieces in $T_v$ and $T_e$ cancel out,
either in $N_{ij}$ or when combining it with $N_{ji}^*$.
Therefore we have\footnote
{Both here and when discussing the coupling of $ch_0$ we assume that $X$ is CY,
so that $q_{123}$ is constant.
We believe the CY condition can be dropped, though details haven't been worked out.}
for $i<j$
\be
\lim_{L \to \infty} \hat a ( \frac{a_j}{a_i}  N_{ji} + \frac{a_i}{a_j}  N_{ij})
=
(-q_{123}^{\frac12})^{|N_{ij}|}
\ee
with
\be
|N_{ij}| = \sum_{v\in\Delta_X^{(0)}} |K_{v,reg,j}|-|K_{v,reg,i}|
 + \sum_{e\in\Delta_X^{(1)}} f_{\lambda_{i,e}} -  f_{\lambda_{j,e}}
\ee
We can write
\be
\prod_{i<j} (-q_{123}^{\frac12})^{|N_{ij}|}
=
\prod_{i=1}^n (-q_{123}^{\frac12})^{(-n-1+2i)(\sum_e f_{\lambda_{i,e}}-\sum_v |K_{i,v,reg}|)}
\ee
This proves factorization along $A_{n-1}$ for any $X$ without D4-branes: summing over fixed points
\be
Z^{7d}_{U(n)} (X) =  \sum_{K} e^{u(K)} \hat a (T)
= \sum_K e^{u} \prod_i \hat a (N_i) \prod_{i<j} (-q_{123}^{\frac12})^{|N_{ij}|}
\ee
where we postpone the discussion of classical parts.

\subsection{Adding faces}

If there are compact 4-cycles, denote fundamental quantities by $\tilde a$, $\tilde K$.
Let $a_i = \tilde a_i q^{m_i}$ and
\be
K_{i,v} =q^{-m_i} \left( \tilde  K_{i,v} - \frac{1-q^{m_i}}{P_{123}} \right)
\ee
where the fluxes $m$ are $n \times n_f$ integers.\footnote
{The full notation is $m_{i,f}$ where $i =1,\ldots,n$ and $f \in \Delta_X^{(2)}$.
Sometimes, we will drop indices.}
The perturbative factor is
\be
\label{pert}
T_{pert} =  -\sum_{v\in \Delta_X^{(0)}} \frac{1}{P^*_{123}} \sum_{i,j} \frac{\tilde a_i}{\tilde a _j}
\ee
The difference of perturbative factors in the two variables
\be
\label{calP}
\mathcal P_m = \sum_{v\in\Delta_X^{(0)}} \frac{1-q^{m}}{P^*_{123}}
\ee
is a Laurent polynomial (so we can take its plethystic) satisfying
$\mathcal P_{-m}=-q_{123}\mathcal P_m^*$.
We have
\be
T = \sum_{i,j} \frac{a_i}{a_j} N_{ij} + \sum_{i,j} \frac{\tilde a_i}{\tilde a_j} \mathcal P_{m_{ij}}
\ee
where $m_{ij} = m_i - m_j$ and we introduced the short notation
\be
q^{m_{ij}} N_{ij} =
\sum_{v\in\Delta_X^{(0)}} q^{m_{ij}} T_{v,ij} + \sum_{v\in\Delta_X^{(0)}} q^{m_{ij}} \sum_\alpha t_{\alpha,ij}
\ee
Summing over fixed points
\be
\widehat Z^{7d}_{U(n)} (X) =  \sum_{K,m} e^{u(K,m)} \hat a (T)
\ee
Looking at \cref{classical}, we observe that now
$\alpha_i = \tilde \alpha_i + m_i \cdot \epsilon$ depends on fixed point data.
The last term in $ch_3$ now contributes as
\be
\sum_{e \in \Delta_X^{(1)}} |\lambda_{i,e}| \sum_{v \in e} \frac{\alpha_i}{\epsilon_e^{(v)}}
=
\sum_{e\in \Delta_X^{(1)}}  \psi \cdot m_i |\lambda_{e,i}|
\label{minch3}
\ee
Hence we get
\be
u^{U(n)}_{K,\lambda,m} (g,t)
= g ch_3 + ch_2 - \frac{n}{24} c_2(X) \cdot t + \frac{ch_0}{g^2 - (n\epsilon/2)^2}
\ee

\subsection{Classicalities}
\label{s:class}

\subsubsection{Details}

Let us explain how to compute
\be
u_0 = \int_X e^{\omega + \ii B} \, ch(F) \wedge \Gamma_X
\ee
At large radius and $B$-field, this expression gives
the central charge of the bound state,
with $ch(F) \wedge \Gamma_X$ being its RR charge.
For our purposes, it's enough to only keep two terms:
\be
\Gamma_X \sim 1 + \frac{\beta^2}{24} c_2(X)
\ee
For non-compact $X$, we define $u_0$ equivariantly.

It is useful to recall that $\mathcal H$ defined in \cref{calH}
encodes $ch(F)$ at the fixed points
\be
\sum_{v \in \Delta_X^{(0)}} P_{123} \mathcal H_v =
\operatorname{tr} e^{\beta \Phi}
\ee
\emph{in the instanton background}, with $\Phi$ the adjoint scalar.
An application of Duistermaat-Heckman theorem then gives
\be
u_0 = \sum_{q=0}^3 \sum_{v \in \Delta_X^{(0)}}
\frac {H_v^q} {\beta^q q!}
\frac{1} {\prod_{a=1}^3 \epsilon_a^{(v)}}
\operatorname{coeff}_{3-q} P_{123} \mathcal H \,
(1 + \frac{\beta^2}{24} \sum_{1 \leq a<b \leq 3} \epsilon_a^{(v)} \epsilon_b^{(v)})
\ee
for any toric threefold, with $\operatorname{coeff}_p$
the coefficient of $\beta^p$ in the small-$\beta$ expansion.
Recalling
\be
P_{123} \mathcal H_v =
\sum_i
\left(
 a_i -P_{123} a_i K_{i,v,reg} - a_i \sum_\alpha \lambda_\alpha \frac{P_{123}}{P_\alpha}
\right)
\ee
and expanding, one arrives at the result \cref{classical}.
Explicitly:
\be
\label{u0}
u_0 = (ch_3 + ch_1 \cdot \Gamma_2)
 + \beta^{-1} (ch_2 + \frac{n}{24} c_2 (X) \cdot t)
 + \beta^{-2} ch_1 + \beta^{-3} ch_0
\ee
where we defined
\be
ch_1 \cdot \Gamma_2 = \frac{1}{24} \sum_{i=1}^n \sum_{v\in\Delta_X^{(0)}} \alpha_i
\sum_{1 \leq a<b \leq 3} \frac {\epsilon_a^{(v)} \epsilon_b^{(v)} }
{\epsilon_1^{(v)} \epsilon_2^{(v)} \epsilon_3^{(v)}}
\ee
and
\be
\label{c2dt}
c_2 (X) \cdot t =
\sum_{v \in\Delta_X^{(0)}}
\frac {H_v} {\epsilon_1^{(v)} \epsilon_2^{(v)}\epsilon_3^{(v)}}
\sum_{1 \leq a<b \leq 3} \epsilon_a^{(v)} \epsilon_b^{(v)}
\ee
Since in the main discussion we are not paying attention to terms linear in $\sum_i m_i$,
the terms $\Gamma_2 \cdot ch_1$ and $ch_1$ have been dropped there.
The same applies to powers of $\beta$,
which are recovered by quantizing $\omega$.

We can write all terms involving only $\alpha$ and $H$ in \cref{classical,u0} as
\be
\label{5dfp}
\sum_{i=1}^n \sum_{v \in\Delta_X^{(0)}}
\sum_{p=0}^3
\frac
{(H_v/\beta)^p \alpha_i^{3-p}}
{p!(3-p)! \epsilon_1^{(v)} \epsilon_2^{(v)}\epsilon_3^{(v)}}
= \sum_{i=1}^n \sum_{v \in\Delta_X^{(0)}} \frac
{(H_v/\beta + \alpha_i)^3}
{3! \epsilon_1^{(v)} \epsilon_2^{(v)}\epsilon_3^{(v)}}
\ee
where $\alpha_i = \tilde \alpha_i + m_i \cdot \epsilon^{(v)}$,
and $m$ can be non-zero only if compact divisors are present.
Luckily, \cref{5dfp} only contributes either terms proportional to powers of $m$,
or terms proportional to powers of $\tilde \alpha$, but not mixed terms.\footnote
{This is a consequence of the fact that
\cref{qdiff} is \emph{a polynomial} in $q$'s \cite{wip},
i.e.~it cannot have singular terms as $\beta \to 0$.}
The former can be computed (see the examples),
while the latter can be discarded as overall constants,
together with the perturbative part in $\tilde a$ variables.
Incidentally, this is the only dependence on 7d Coulomb moduli left
if we trust our working assumption.

Finally, we can turn on flat RR potentials $C_{p+1}^{RR}$ ($p$ even)
that couple to the RR charge, thus promoting $u_0$ to
\be
\label{u}
u = \int_{S^1 \times X}
\left( \frac{ds}{g_s} e^{\omega + \ii B} + \ii \sum_p C_{p+1}^{RR} \right)
\, ch(F) \wedge \Gamma_X
\ee
For $D0$-branes, this gives the complexified (dimensionless) quantity
\be
g := \frac{\beta}{\ell_s g_s} + \ii \int_{S^1} C_1^{RR}
\ee
which multiplies $ch_3$ in \cref{our-def-classical},
with $R \sim \ell_s g_s$ the TN-radius.
For higher $Dp$-branes, it is unclear how to extend equivariantly $C^{RR}$
from first principles, so as to make \cref{u} well-defined.
However, \cref{5dfp} suggests a 5d fixed-point interpretation:
the index $i$ runs over 2d fixed points,
and index $v$ over 3d ones.
If we replace our $\alpha_i$ with the 2d Hamiltonian $H_i$ on $TN_n$ space,
and weigh it by the corresponding tangent weights $\epsilon_{4,5}^{(i)}$ at the fixed point,
\be
\sum_{i,v} \frac {(H_v + \alpha_i)^3}
{3! \epsilon_1^{(v)} \epsilon_2^{(v)} \epsilon_3^{(v)} }
\to
\sum_{i,v} \frac {(H_v + H_i)^3}
{3! \epsilon_1^{(v)} \epsilon_2^{(v)} \epsilon_3^{(v)} \epsilon_4^{(i)} \epsilon_5^{(i)} }
\ee
then we can read off the remaining coupling
(the coupling for $ch_0$ in \cref{our-def-classical})
from \cref{volAn}:
just interpret the $\alpha_i$'s of $A_n$ space as the $\alpha_i$'s of our gauge theory,
neglecting the cubic adjoint terms
(which will pop up again as the limit of the perturbative part).
The obtained action \cref{our-def-classical} is a suitable candidate
for the equivariant extension of \cref{u},
and it satisfies many non-trivial checks (see below),
in particular factorizability.

\subsubsection{Shift equations}

The term $ch_0$ in \cref{classical} is problematic for non-compact $X$.
Let us define
\be
\label{calFr}
\mathcal F (t, \varepsilon) = -
\sum_{v \in \Delta_X^{(0)}} \frac{H^3_v}{3! \epsilon_1^{(v)} \epsilon_2^{(v)} \epsilon_3^{(v)}}
\ee
with notations as in \cref{toric-review}.
This is a regularized triple intersection for $X$.
We have evidence \cite{wip} that,
when $X$ has at least one compact four-cycle,
\be
- \sum_{v \in\Delta_X^{(0)}} \frac
{(H_v + \epsilon\cdot m)^3}
{3! \epsilon_1^{(v)} \epsilon_2^{(v)}\epsilon_3^{(v)}}
=
\mathcal F(t,\varepsilon) +
\mathcal F_{shift}(t,m)
\ee
where we used the way $\alpha$ is shifted in \cref{5dfp}
and $\mathcal F_{shift}$ is a function of $t,m$ independent of regulators.
If we choose\footnote
{This corresponds to setting $\varepsilon$ to zero for compact divisors.
We do not need to make this specific choice and one can perform the analysis in general.
However, the proper geometric treatment of this problem is
beyond the scope of this work and will be explained elsewhere \cite{wip}.}
some of the $\varepsilon$ such that
\be
- \sum_{v \in\Delta_X^{(0)}} \frac
{(H_v + \epsilon\cdot m)^3}
{3! \epsilon_1^{(v)} \epsilon_2^{(v)}\epsilon_3^{(v)}}
=
\mathcal F(t -\psi \cdot m,\varepsilon)
\ee
then we have
\be
 \mathcal F(t,\varepsilon)= \mathcal F(t-\psi \cdot m,\varepsilon) - \mathcal F_{shift}(t,m)
\ee
By also choosing $m$ such that $t-\psi\cdot m=0$
(choosing $\dim H_4$ out of $\dim H_2$ $t$ variables),
we get a prescription to compute the regularized triple intersection as
$-\mathcal F_{shift}(t,m)$ in terms of DH sums.

Likewise, if in \cref{c2dt} we set $\varepsilon$'s corresponding to compact divisors to zero,
we can study the difference $c_2(X) \cdot (t + \psi\cdot m) - c_2(X) \cdot t$.

We spell this out for some examples in \cref{s:examples}.

\subsection{7d master formula}
\label{7d-master-sec}

Let us enforce our Coulomb independence assumption.
Setting $\tilde a_i = L^i$ and taking the large $L$ limit with $i<j$, we get
(this equality is proved momentarily in \cref{proof})
\be
\label{key-sign}
\lim_{L \to \infty} \hat a
\left(
\frac{a_j}{a_i}  N_{ji} + \frac{a_i}{a_j}  N_{ij}
\right)
=
(-q_{123}^{\frac12})^{s_{ij}}
\ee
with the integer given by
\be
s_{ij} =
|N_{ij}| +
\sum_{e\in\Delta_X^{(1)}} (|\lambda_{i,e}| + |\lambda_{j,e}|) (\psi \cdot m_{ij})
\ee
This proves factorization for $\hat a(N_i)$.
Summing over fixed points, we get
\be
\widehat Z = \sum_{m,K} e^u \prod_{i<j} \left( -q_{123}^{\frac12}\right) ^{s_{ij}+|\mathcal P_{m_{ij}}|}
\prod_{i=1}^n \hat a (N_i)
\ee
Let $m_* = \sum_{i=1}^n m_i$ and
\be
\sigma_\ell(m) := \sum_{i=1}^{\ell-1} m_i - \sum^n_{i=\ell+1} m_i =
2\sum_{i=1}^\ell m_i-m_\ell-\sum_{i=1}^n m_i\label{def-sigma}
\ee
\be
g_i = g + \frac\epsilon2(n+1-2i)
\label{gi}
\ee

Using results from \cref{app:combin} and some extra tools \cite{wip}, one shows that
\be
\label{clas-fact}
u^{U(n)}_{K,\lambda,m} (g,t) + \frac{\epsilon}2 \sum_{i<j} (s_{ij} + |\mathcal P_{m_{ij}}|)
=
\sum_i u^{U(1)}_{K_i,\lambda_i,0}
\left( g_i,t + g_i \psi \cdot m_i+ \frac{\epsilon}2 \psi\cdot \sigma_i \right)
\mod{m_*}
\ee
and arrives at the 7d master formula 
\be
\label{7d-master}
\boxed{
\widehat Z^{7d}_{U(n)} (X; p,Q_e) =
\sum_{m \in \BZ^{n \times n_f}}
e^{f(m_*)}
\prod_{i=1}^n
\sum_{K_i,\lambda_i}
e^{u^{U(1)}_{K_i,\lambda_i,0} \left( g_i,t + g_i \psi \cdot m_i+ \frac{\epsilon}2 \psi\cdot \sigma_i \right)}
\,
\hat a (N_i)
}
\ee
where the partition function completely factorizes in each $m_*$ sector.
The function of $f(m_*)$,
which is computable in our formalism up to a term linear in $m_*$ coming from $c_1$,
is a cubic polynomial in $m_*$ that goes to zero for $m_*=0$.


If we normalize by the empty vertex and use \cref{swear},
we can match \emph{exactly} the 5d gauge theory result \cref{5d-master} upon setting $m_*=0$,
for any geometry $X$ engineering theory $\TX$, if in \cref{ecy,An-local-global} we identify
\be
\epsilon_4 + \epsilon_5 = \epsilon,
\qquad
n \frac{\epsilon_4-\epsilon_5}2 = g
\ee
It is amusing to observe that in these cases $n_f = \operatorname{rk} \TX$.

The map between $m_i$ and $h_i$ in \cref{5d-master} depends on the details of geometric engineering.
Until now we denoted the dependence of $Z^{7d}$ on $t_e$, $e \in \Delta_X^{(1)}$
in order to have a clear interpretation of the various shifts.
Before applying the geometric engineering dictionary and as discussed in \cref{toric-review},
we expand $t_e$ in a basis of $H_2(X)$.

\subsubsection{Proof}
\label{proof}

Let us prove \cref{key-sign}.
Setting $\tilde a_i = L^i$ we compute for  $i<j$
\be
\lim_{L \to \infty} \hat a (\frac{a_j}{a_i}  N_{ji} + \frac{a_i}{a_j}  N_{ij})
\ee
The first observation is that this is equal to
\be
(-q_{123}^{\frac12})^{|q^{m_{ij}} N_{ij}|}
\ee
So we just need to compute its net size.
The first term gives
\be
\sum_{v \in \Delta_X^{(0)}} |K_{j,v,reg}| - |K_{i,v,reg}|
\ee
The second term  gets a contribution from $\lambda^*_j$ and one from $-q_{123} \lambda_i$.
The first one gives
\be
\lim_{q_1 \to 1}
\sum_{(a,b) \in \lambda_{e,j}}
\frac{-1 + q_1^{-1-\psi \cdot m_{ij} +\psi_2(a-1)+\psi_3(b-1)}}{1-q_1}
=
\psi \cdot m_{ij} |\lambda_{e,j}| - f_{\lambda_{e,j}}
\ee
where in intermediate steps we can take the edge along direction 1.
The second one gives
\be
\psi \cdot m_{ij} |\lambda_{e,i}| + f_{\lambda_{e,i}}
\ee
Combining them we get the result.

\section{Experimental evidence}
\label{s:examples}
We discuss some examples in detail, focusing on some of the simplest cases.
Many more examples could be added.
Our purpose here is to explain in detail notations and perform explicit checks of general results.

\subsection{\texorpdfstring{$SU(N)$}{SU(N)} examples}

With notations as in \cite{Closset:2018bjz},
the 5d SCFT giving the UV completion of 5d $\CalN=1$ $SU(N)_k$ gauge theory
is obtained in M-theory on a singularity whose toric diagram has external points at
\be
\label{eq:toricSUNk}
D_0 = (0, 0)\,, \quad D_N= (0,N)\,, \quad D_x= (-1, w_x)\,, \quad D_y=(1, w_y)\,,
\ee
with $w_x, w_y \in \BZ$.
We impose the convexity condition
\be
\label{eq:convexSUN}
0<w< 2N, \quad w\equiv w_x + w_y
\ee
The Chern-Simons level is $k = w - N$.
The toric divisors satisfy relations
\be
\begin{aligned}
D_0 &\cong  (N-1) D_N + (w-2)D_x+\sum_{a=1}^{N-1} (a-1) E_a\\
D_N &\cong - D_0 - 2 D_x - \sum_{a=1}^{N-1} E_a\\
D_x &\cong D_y.
\end{aligned}
\ee
\begin{figure}[htb]
\begin{center}
\includegraphics[scale=0.5]{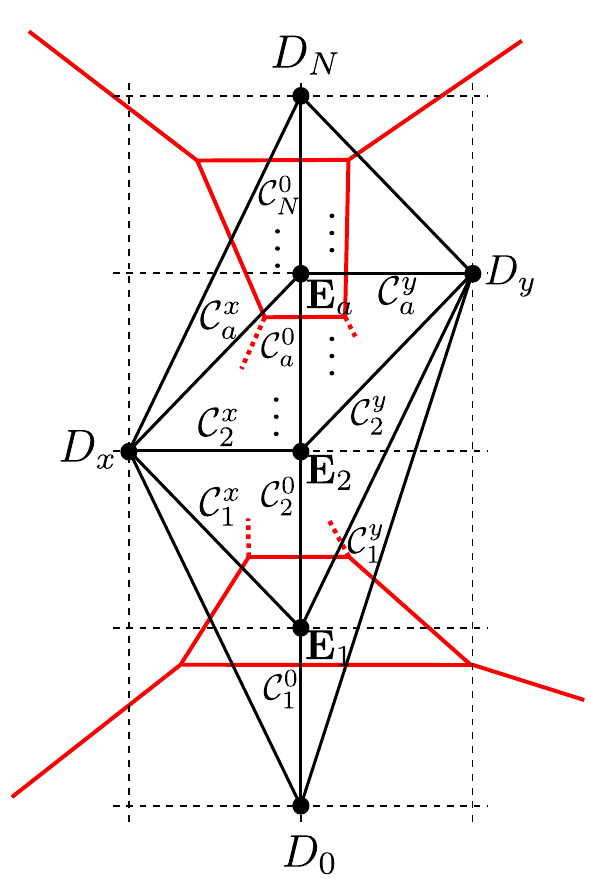}
\end{center}
\caption{The $SU(N)_k$ gauge theory phase.
We denote by $D_0$, $D_N$, $D_x$ and $D_y$ the non-compact divisors
corresponding to external points in \cref{eq:toricSUNk}
and $E_1, \cdots, E_{N-1}$ the compact divisors,
corresponding to internal points $(0, a)$ for $a=1,\ldots,N-1$.}
\label{fig:SUNkgeo}
\end{figure}
The resolution in \cref{fig:SUNkgeo} contains the curves
\begin{multline}
C_a^x \cong D_x \cdot E_a~, \quad C_a^y \cong D_y \cdot E_a~, \quad a=1, \cdots N-1~,\\
C_a^0 \cong E_{a-1} \cdot E_a~, \quad a= 1, \cdots N~,
\end{multline}
where we denoted $D_0$ as $E_0$ and $D_N$ as $E_N$.
One finds
\be
\label{eq:edgequation}
C_a^0 - C_{a+1}^0 \cong (w-2 a) C_a^x~, \quad a=1, \cdots, N-1~.
\ee
One can intersect the $N$ independent curves $(C_1^0, C_a^x)$,
whose volumes are $vol (C_1^0)=t_1$, $vol (C_a^x) = t_{a+1}$,
with divisors $(D_0, E_b, D_N, D_x, D_y)$ to get the GLSM description
\be
Q=
\begin{pmatrix}
 w-2 & -w \delta_{1,b} & 0 & 1& 1 \\
 \delta_{a,1}& -A_{ab} & \delta_{a, N-1} & 0& 0
\end{pmatrix}
\ee
with $a, b = 1, \cdots, N-1$, and
\be
A_{ab} = 2 \delta_{ab} - \delta_{a,b+1}-\delta_{a+1,b}.
\ee
With notations as in \cref{toric-review},
the toric variety obtained from symplectic quotient engineers
$SU(N)$ gauge theory with Chern-Simons level $k$.
We can define
\be
J = \mu_x D_x + \sum_{a=1}^{N-1} \nu_a E_a~.
\ee
The parameters $\nu_a = - \varphi_a$, $\mu_x = h$  are related to the FI parameters by
\be
\label{eq:generalkalerna}
t_1 = h +  (k+N) \varphi_1~, \quad t_{a+1} = \sum_{b} A_{ab} \varphi_b
\ee
Taking the cube one finds the field theory prepotential for $SU(N)_k$,
\be
\mathcal F = - \frac16  J^3 =
- \frac12 \mu_x^2 \nu_a (D_x^2 E^a) - \frac12 \mu_x \nu_a \nu_b (D_x E^a E^b)
- \frac16 \nu_a \nu_b \nu_c (E^aE^bE^c)
\ee
where we set\footnote
{Ref.~\cite{Foscolo:2017vzf} allows to extend
the usual tools of Hodge theory to the non-compact CY3 setting.}
$D_x^3=0$.
The non-zero triple-intersections are
\be
D_x E_a E_b = - A_{ab}~, \quad E_a^3=8~, \quad E_{a-1}^2 E_a = w-2 a~, \quad
E_{a-1} E_a^2 = 2 a-2 -w~.
\ee

\subsection{The case of \texorpdfstring{$SU(2)_0$}{SU2 lev 0}}

We consider the CY manifold $X=\CalO (-2,-2) \to \BP^1\times \BP^1$,
which corresponds to the 5d theory with $SU(2)$ gauge group and zero Chern-Simons level.

The toric variety $X$ can be constructed as the K\"ahler
quotient of $\BC^5$ by $U(1)^2$ with the action defined by the charge matrix
\be
Q =
\begin{pmatrix}
1 & 1 & 0 & 0 & -2 \\
0 & 0 & 1 & 1 & -2
\end{pmatrix}~,
\ee
and moment maps
\be
\begin{aligned}
|z_1|^2 + |z_2|^2 - 2 |z_5|^2 &= t_1~,\\
|z_3|^2 + |z_4|^2 - 2 |z_5|^2 &= t_2~,
\end{aligned}
\ee
where we have assumed that $\{z_i\}, i=1,\ldots,5$ parametrize $\BC^5$.
This toric manifold can be covered by 4 affine charts associated
to the fixed points under the $T^3$-action.
These charts can be parametrized with the set of coordinates
that we summarize below in the table:
\be
\begin{tabu}{c|c|c|c}
	\text{vtx} & \text{gauge invt coords} & \Omega & H\\
\hline
	1 & z_2/z_1, z_4/z_3, z_5 z_1^2 z_3^2 &
	\varepsilon_2-\varepsilon_1, \varepsilon_4-\varepsilon_3,\varepsilon_5+2\varepsilon_1+2\varepsilon_3 &
	\varepsilon_1 t_1+\varepsilon_3 t_2 \\
	2 & z_2/z_1, z_3/z_4, z_5 z_1^2 z_4^2 &
	\varepsilon_2-\varepsilon_1, \varepsilon_3-\varepsilon_4,\varepsilon_5+2\varepsilon_1+2\varepsilon_4 &
	\varepsilon_1 t_1+\varepsilon_4 t_2 \\
	3 & z_1/z_2, z_4/z_3, z_5 z_2^2 z_3^2 &
	\varepsilon_1-\varepsilon_2, \varepsilon_4-\varepsilon_3,\varepsilon_5+2\varepsilon_2+2\varepsilon_3 &
	\varepsilon_2 t_1+\varepsilon_3 t_2 \\
	4 & z_1/z_2, z_3/z_4, z_5 z_2^2 z_4^2 &
	\varepsilon_1-\varepsilon_2, \varepsilon_3-\varepsilon_4,\varepsilon_5+2\varepsilon_2+2\varepsilon_4 &
	\varepsilon_2 t_1+\varepsilon_4 t_2
\end{tabu}
\label{table1-SU(2)}
\ee
where the third column corresponds to the $T^3$-action
at the corresponding fixed point written in terms of $\varepsilon_i$
that parametrize $T^5$ acting on $\BC^5$.
The last column corresponds to the value of the Hamiltonian
$H= \sum_{i=1}^5 \varepsilon_i |z_i|^2$ at each fixed point.
Alternatively we can parametrize the $T^3$-action
in terms of three independent (global) $(\epsilon_1, \epsilon_2, \epsilon_3)$
\be
\begin{tabu}{c|l}
\text{vtx} & \Omega \\
\hline
	1 &   \epsilon_1 , \epsilon_2, \epsilon_3   \\
	2 &   \epsilon_1, -\epsilon_2,  \epsilon_3 + 2 \epsilon_2  \\
	3 &    - \epsilon_1,  \epsilon_2,   \epsilon_3 + 2\epsilon_1   \\
	4 &  -\epsilon_1, - \epsilon_2,  \epsilon_3 + 2 \epsilon_1 + 2 \epsilon_2
\end{tabu}
\label{defin-eps-SU(2)}
\ee
If we denote by $H_v$ ($v=1,2,3,4$) the Hamiltonian at the fixed points we have
\be
\begin{aligned}
 H_2 - H_1 = \epsilon_2 t_2 \\
 H_3 - H_1 = \epsilon_1 t_1 \\
 H_4 - H_3 = \epsilon_2 t_2 \\
 H_4 - H_2 = \epsilon_1 t_1
\end{aligned}
\label{dif-H-SU(2)}
\ee
which are expressed in terms of global $(\epsilon_1, \epsilon_2, \epsilon_3)$.
These shifts are uniquely fixed by the compact $\BP^1$'s.
The relevant geometry (vertices and edges) is conveniently summarized by
\be
\begin{tikzpicture}
  \matrix (m) [matrix of math nodes,row sep=3em,column sep=8em,minimum width=2em]
  {
     v_1 & v_2 \\
     v_3 & v_4 \\};
  \path[-]
	  (m-1-1) edge node [left] {$-2$} node [right] {0} (m-2-1)
    edge node [above] {$-2$} node [below] {0} (m-1-2)
    (m-1-2) edge node [right] {$-2$} node [left] {0} (m-2-2)
    (m-2-1) edge node [above] {0} node [below] {$-2$} (m-2-2);
\end{tikzpicture}
\qquad
\begin{tikzpicture}
  \matrix (m) [matrix of math nodes,row sep=3em,column sep=8em,minimum width=2em]
  {
     v_1 & v_2 \\
     v_3 & v_4 \\};
  \path[-]
    (m-1-1) edge node [left] {$t_1$} (m-2-1)
    edge node [above] {$t_2$}  (m-1-2)
    (m-1-2) edge node [right] {$t_1$} (m-2-2)
    (m-2-1) edge node [below] {$t_2$} (m-2-2);
\end{tikzpicture}
\ee
where the first diagram keeps track of $\psi$ data, the second of edge sizes $t_e$.
The geometry has one compact face, so $m \in \BZ^n$,
and \cref{edge-face} becomes for all four edges
\be
\psi \cdot m_\ell = -2 m_\ell~.
\ee

Using this toric data we can perform the explicit calculations
relevant for 7d theory on this geometry.
The contribution of fluxes to the classical terms in \cref{classical} is
\be
\label{chmsu2}
\begin{aligned}
- \sum_i \sum_{v \in \Delta_X^{(0)}} \frac{(\tilde\alpha_i + m_i \epsilon_3^{(v)})^3}{3! \epsilon_1^{(v)} \epsilon_2^{(v)}\epsilon_3^{(v)}}
&= - \sum_i \sum_{v \in \Delta_X^{(0)}} \frac{\tilde\alpha_i^3}{3! \epsilon_1^{(v)} \epsilon_2^{(v)}\epsilon_3^{(v)}} - \frac{4}{3} \sum_i m_i^3~, \\
- \sum_i \sum_{v\in\Delta_X^{(0)}} \frac{(\tilde\alpha_i+ m_i \epsilon_3^{(v)})^2 H_v}{2 \epsilon_1^{(v)} \epsilon_2^{(v)}\epsilon_3^{(v)}}
&= - \sum_i \sum_{v\in\Delta_X^{(0)}} \frac{\tilde\alpha_i^2 H_v}{2 \epsilon_1^{(v)} \epsilon_2^{(v)}\epsilon_3^{(v)}} - (t_1 + t_2) \sum_i m_i^2~, \\
-\sum_i \sum_{v\in\Delta_X^{(0)}} \frac{(\tilde\alpha_i + m_i \epsilon_3^{(v)})H^2_v}{2 \epsilon_1^{(v)} \epsilon_2^{(v)}\epsilon_3^{(v)}}
&= -\sum_i \sum_{v\in\Delta_X^{(0)}} \frac{\tilde\alpha_i H^2_v}{2 \epsilon_1^{(v)} \epsilon_2^{(v)}\epsilon_3^{(v)}} - t_1 t_2 \sum_i m_i~,
\end{aligned}
\ee
where we used \cref{defin-eps-SU(2),dif-H-SU(2)}.
The classical action (with $\tilde \alpha=0$) is built out of
\be
\begin{aligned}
ch_3 &=
-\frac43 \sum_i m_i^3 + \sum_i \left( \sum_v |K_{i,v}^{reg}| - \sum_e f_{\lambda_{e,i}} \right) - \sum_{i,e} \psi\cdot m_i |\lambda_{e,i}|~, \\
ch_2 &= -(t_1+t_2) \sum_i m_i^2 -\sum_{i,e} t_e |\lambda_{i,e}| ~,\\
ch_1 &= -t_1 t_2 \sum_i m_i ~,\\
ch_0 &= n \mathcal F(t)~,
\end{aligned}
\ee
where the last term requires a separate discussion.
In \cref{calFr} we define ${\cal F}(t, \varepsilon)$.
Using the explicit toric data from  \cref{table1-SU(2)}
and setting $\varepsilon_5=0$ in \cref{calFr} we get
\be
 {\mathcal F}(t_1, t_2 , \varepsilon) =
 \frac{1}{12} t_2^2 (-3 t_1 + t_2) + f(\varepsilon) (t_1 - t_2)^3~,\label{defFe-SU(2)}
\ee
where
\be
 f (\varepsilon)=
 \frac
   { \varepsilon_2^2 \varepsilon_3 \varepsilon_4 + \varepsilon_1^2 (\varepsilon_2 + \varepsilon_3) (\varepsilon_2 + \varepsilon_4) 
   + \varepsilon_1 \varepsilon_2 (\varepsilon_3 \varepsilon_4 + \varepsilon_2 (\varepsilon_3 + \varepsilon_4))}{12 (\varepsilon_1 + 
   \varepsilon_3) (\varepsilon_2 + \varepsilon_3) (\varepsilon_1 + \varepsilon_4) (\varepsilon_2 + \varepsilon_4)}~.
\ee
The ${\mathcal F}(t_1, t_2 , \varepsilon)$ has the property
\be
 {\mathcal F} (t_1 + 2m, t_2 + 2m, \varepsilon) =
 {\mathcal F} (t_1, t_2,\varepsilon) - m t_1 t_2 - m^2 (t_1 + t_2) - \frac{4}{3} m^3~,
 \label{shift-SU(2)}
\ee
where terms in $m$ coincide with terms from \cref{chmsu2}.
We extract the universal part
\be
\mathcal F (t_1,t_2)  = \frac{1}{12} t_2^2 (-3 t_1 + t_2)~,\label{canF-SU(2)}
\ee
but we stress that we can also use ${\mathcal F}(t_1, t_2 , \varepsilon)$
from \cref{defFe-SU(2)}
since in what follows we only use the shift symmetry \cref{shift-SU(2)}.


Finally let us compute the polynomial $\mathcal P$, defined in \cref{calP}, for this example:
\begin{multline}
q_{123}^* \mathcal  P_m (q_1, q_2, q_3)
= \frac{q_3^m -1}{(1-q_1)(1-q_2)(1-q_3)} + \frac{q_3^m  q_2^{2m}-1}{(1-q_1)(1-q^{-1}_2)(1-q_3q_2^2)} \\
+ \frac{q_3^m  q_1^{2m}-1}{(1-q^{-1}_1)(1-q_2)(1-q_3q_1^2)} + \frac{q_3^m q_1^{2m} q_2^{2m}-1}{(1-q^{-1}_1)(1-q^{-1}_2)(1-q_3 q_1^2 q_2^2)}~.
\end{multline}
For $m>0$ we get
\be
 q_{123}^* \mathcal  P_m (q_1, q_2, q_3) =
 - \sum_{s=0}^{m-1} \sum_{l=0}^{2s} \sum_{k=0}^{2s} q_3^s q_1^l q_2^k~.
\ee
Using the standard identities
\be
 \sum_{s=1}^n s = \frac{n(n+1)}{2},\quad \sum_{s=1}^{n} s^2 = \frac{n(n+1) (2n+1)}{6}
\ee
we get
\be
   |\mathcal P_m| = \mathcal P_m (1, 1, 1) = - \sum_{s=0}^{m-1} (2s+1)^2
   = \frac{1}{3} \left( m- 4 m^3 \right )~,
\ee
which is an integer, as expected.
For $m<0$ we use the property
\be
 \mathcal  P_{-m} (q_1, q_2, q_3) = - q^{-1}_{123} \mathcal P_m (q^{-1}_1, q^{-1}_2, q^{-1}_3)~,
\ee
which implies
\be
 |\mathcal P_{-m}| = - |\mathcal P_m|~,
\ee
so it is clear that $|\mathcal P_m|$ is an odd function of $m$.

Using identities from \cref{app:combin}, we have
\begin{multline}
\sum_{i=1}^n
\frac {\mathcal F(t + \psi\cdot g_i m_i+\frac{\epsilon}2 \psi\cdot\sigma_i)} {g_i^2 - (\epsilon/2)^2}
=
\frac43 \left( \frac{\epsilon}2 \sum_{i<j} m_{ij}^3
+\frac{\epsilon^2 g}{4g^2-(n\epsilon)^2} m_*^3
+g \sum_i m_i^3 \right)
+\frac{n \mathcal F(t)} {g^2 - (n \epsilon/2)^2} \\
\quad
-(t_1+t_2) \left(\frac{n \epsilon^2}{4g^2-(n\epsilon)^2} m_*^2
+\sum_i m_i^2 \right)
+t_1t_2 \frac{g m_*} {g^2 - (n \epsilon/2)^2} \label{SU2-classical-dec}
\end{multline}
Alternatively, we can write it as
\begin{multline}
\sum_{i=1}^n
\frac {\mathcal F(t + \psi\cdot g_i m_i+\frac{\epsilon}2 \psi\cdot\sigma_i)} {g_i^2 - (\epsilon/2)^2}
= \frac{\epsilon}2 \frac43 \sum_{i<j} m_{ij}^3
 + \frac{n \mathcal F(t + \psi \cdot g \frac{m_*}{n})} {g^2 - (n \epsilon/2)^2}
 + \frac43 g \sum_i \left( m_i - \frac{m_*}{n} \right )^3 \\
 - \left ( (t_1 + \psi \cdot g \frac{m_*}{n}) + (t_2 + \psi \cdot g \frac{m_*}{n}) \right )
 \sum_i \left( m_i - \frac{m_*}{n} \right )^2~.
\label{SU2-alternative-rel}
\end{multline}
The first term in RHS of \cref{SU2-classical-dec} comes from $|\mathcal P_m|=\frac13(m-4 m^3)$,
while the other term in $\mathcal P$ combines with $c_2(X)$.
Indeed, using our prescription \cref{c2dt} for $c_2(X)\cdot t$,
we can write the factorization formulas for the classical action:
up to terms proportional to $m_*$, we get
\begin{multline}
u + \frac{\epsilon}2 \sum_{i<j} (s_{ij} + |\mathcal P_{m_{ij}}|)
=
\sum_i \frac{\mathcal F(t + g_i \psi \cdot m_i
+\frac\epsilon2 \psi\cdot\sigma_i)}{g^2_i - (\epsilon/2)^2} \\
+\sum_i \left[
g_i \left( \sum_v |K_{i,v}^{reg}|-\sum_e f_{\lambda_{i,e}} \right)
- \sum_{e} |\lambda_{i,e}| (t_e + g_i \psi \cdot m_i+\psi\cdot \sigma_i \frac{\epsilon}2)
	\right] \\
- \frac{1}{24} c_2(X)\cdot (nt + 2 \frac\epsilon2 \sum_{i<j} \psi \cdot m_{ij})
\end{multline}
in agreement with \cref{clas-fact}.
We use the property
\be
-\frac{1}{24} c_2(X) \cdot (t + \psi \cdot m) + \frac{1}{24} c_2(X) \cdot t
= \frac{m}{6}
\ee
which can be checked explicitly from \cref{c2dt}.

Finally we use the geometric engineering dictionary for $SU(2)$ theory
where K\"ahler parameters $(t_1, t_2)$ are related
to the scalar $\varphi$ in $SU(2)$ vector multiplet and the coupling $h$ as
\be
t_1 = h + 2\varphi,
\quad
t_2 = 2 \varphi~.
\ee
We then match \emph{exactly} the 7d and 5d master formulas \cref{7d-master,5d-master}
by identifying
\be
h_{\ell} = -\sum_{i=1}^\ell m_i + \frac12 m_* = \frac12 \left( -m_1-\cdots-m_\ell+m_{\ell+1} + \cdots + m_n \right)
\ee
and imposing the condition $m_* = 0$,
which implies $h_0=h_n=0$ and amounts to going from $U(n)$ to $SU(n)$ in 7d.
We conclude that the partition function for
the 7d $SU(n)$ theory on $X=\CalO (-2,-2) \to \BP^1\times \BP^1$
is the same as the partition function for
the 5d $SU(2)_0$ theory on $A_{n-1}$ space
(both theories are extended to $S^1$ in the appropriate fashion).
The classical part \cref{canF-SU(2)} becomes
\be
 {\cal F} (\phi, h) = - h \varphi^2 - \frac{4}{3} \varphi^3~,
\ee
as it should be.
If instead we use \cref{defFe-SU(2)}, then we have
\be
 {\cal F} (\phi, h) = - h \varphi^2 - \frac{4}{3} \varphi^3 + f(\varepsilon) h^3~,
\ee
which may correspond to adding some non-dynamical (purely geometric) term
on the 5d side.

\subsection{A rank two example: \texorpdfstring{$SU(3)_0$}{SU3 lev 0}}

Next we consider another example of CY that corresponds to $SU(3)$ 5d gauge theory with zero Chern-Simons level. This CY can be obtained by 
 the K\"aher quotient of $\BC^6$ by $U(1)^3$ with action defined by the charge matrix
\be
Q =
\begin{pmatrix}
1 & 1 & 1  & -3 &  0  &  0  \\
 0 & 0  & 1  & -2 & 1  & 0   \\
 0 &  0 & 0   &  1 & -2 & 1  
\end{pmatrix}
\ee
 and moment maps 
\be
\begin{aligned}
 |z_1|^2 + |z_2|^2 + |z_3|^2 - 3 |z_4|^2 = t_1\\
|z_3|^2 - 2|z_4|^2 + |z_5|^2 = t_2\\
 |z_4|^2 - 2 |z_5|^2 + |z_6|^2 = t_3
\end{aligned}
\ee
where we use $\BC^6$ coordinates.
The resulting manifold can be covered by 6 affine chats associated to the fixed points of $T^3$ action. 
 We summarize this in the following tables:
\be
\begin{tabu}{c|c|c}
	\text{vtx} & \text{gauge invt coords} & H\\
\hline
1 & z_2^3 z_4 z_5^2 z_6^3, z_3 z_2^{-1} z_5^{-1} z_6^{-2}, z_1 z_2^{-1} &
\varepsilon_2 t_1+\varepsilon_5t_2+\varepsilon_6(t_3+2t_2) \\
2 & z_1^3 z_4 z_5^2 z_6^3, z_3 z_1^{-1} z_5^{-1} z_6^{-2}, z_2 z_1^{-1} &
\varepsilon_1 t_1+\varepsilon_5t_2+\varepsilon_6(t_3+2t_2) \\
3 & z_2 z_5 z_6^2 z_3^{-1}, z_3^2 z_4 z_2 z_6^{-1}, z_1 z_2^{-1} &
\varepsilon_2 (t_1-t_2)+\varepsilon_3t_2+\varepsilon_6t_3 \\
4 & z_1 z_5 z_6^2 z_3^{-1}, z_3^2 z_4 z_1 z_6^{-1}, z_2 z_1^{-1} &
\varepsilon_1 (t_1-t_2)+\varepsilon_3t_2+\varepsilon_6t_3 \\
5 & z_6 z_3^{-2} z_4^{-1} z_2^{-1}, z_5 z_4^{2} z_3^{3} z_2^{3}, z_1 z_2^{-1} &
\varepsilon_2 (t_3+t_1-t_2)+\varepsilon_3(2t_3+t_2)+\varepsilon_4t_3 \\
6 & z_6 z_3^{-2} z_4^{-1} z_1^{-1}, z_5 z_4^{2} z_3^{3} z_1^{3}, z_2 z_1^{-1} &
\varepsilon_1 (t_3+t_1-t_2)+\varepsilon_3(2t_3+t_2)+\varepsilon_4t_3
\end{tabu}
\ee
 where in middle column we define the coordinates in every chart and in the right column we write the value of Hamiltonian $H=\sum_{i=1}^6 \varepsilon_i |z_i|^2$
  at the corresponding fixed point. The $T^3$ action at every fixed point can be summarized in the following table
\be
\begin{tabu}{c|c}
	\text{vtx} & \Omega \\
\hline
1 &  3\varepsilon_2 +\varepsilon_4 + 2 \varepsilon_5 + 3 \varepsilon_6, \varepsilon_3 -\varepsilon_2 -\varepsilon_5 - 2\varepsilon_6, \varepsilon_1 - \varepsilon_2\\
2 & 3\varepsilon_1 +\varepsilon_4 + 2 \varepsilon_5 + 3 \varepsilon_6, \varepsilon_3 -\varepsilon_1 -\varepsilon_5 - 2\varepsilon_6, \varepsilon_2 - \varepsilon_1\\
3 & \varepsilon_2 + \varepsilon_5 + 2\varepsilon_6 - \varepsilon_3, 2\varepsilon_3 + \varepsilon_4 +\varepsilon_2 - \varepsilon_6, \varepsilon_1-\varepsilon_2\\
4 & \varepsilon_1+\varepsilon_5 +2\varepsilon_6 -\varepsilon_3, 2\varepsilon_3 +\varepsilon_4 + \varepsilon_1 - \varepsilon_6, \varepsilon_2 - \varepsilon_1\\
5 &  \varepsilon_6- 2\varepsilon_3 -\varepsilon_4 -\varepsilon_2, \varepsilon_5 + 2\varepsilon_4 +3\varepsilon_3 +3\varepsilon_2, \varepsilon_1 - \varepsilon_2\\ 
6 & \varepsilon_6 - 2\varepsilon_3 -\varepsilon_4 -\varepsilon_1, \varepsilon_5 +2\varepsilon_4 +3\varepsilon_3 +3\varepsilon_1, \varepsilon_2 - \varepsilon_1
\end{tabu}
\ee
 where we use $\BC^6$ parameters. Equivalently we can rewrite it in terms of 3-independent parameters $(\epsilon_1, \epsilon_2, \epsilon_3)$
\be
\begin{tabu}{c|l}
\text{vtx} & \text{gauge invt coords} \\
\hline
1 &  \epsilon_1, \epsilon_2,  \epsilon_3 \\
2 &  3 \epsilon_3 + \epsilon_1,   \epsilon_2 - \epsilon_3, - \epsilon_3\\
3 &  -\epsilon_2, \epsilon_1 +2 \epsilon_2,   \epsilon_3 \\
4 &  \epsilon_3-\epsilon_2,  2\epsilon_2 + \epsilon_3 + \epsilon_1,  -\epsilon_3 \\
5 &  -\epsilon_1 -2\epsilon_2,   2\epsilon_1 + 3\epsilon_2,  \epsilon_3   \\
6 &  -2\epsilon_2 - \epsilon_3 -\epsilon_1,   2\epsilon_1 + 3\epsilon_3 + 3\epsilon_2,  - \epsilon_3
\end{tabu}
\label{table-gle-SU(3)}
\ee
 If we denote by $H_v$ the value of the Hamiltonian at the fixed point $v$ then the difference of Hamiltonians reads
\be
\begin{aligned}
H_2 - H_1 &= \epsilon_3 t_1 \\
H_3 - H_1 &= \epsilon_2 t_2 \\
H_4 - H_2 &= (\epsilon_2 - \epsilon_3) t_2 \\
H_4 - H_3 &= \epsilon_3 (t_1 - t_2) \\
H_5 - H_3 &= (\epsilon_1 + 2 \epsilon_2) t_3 \\
H_6 - H_5 &= \epsilon_3 (t_3 + t_1 - t_2) \\
H_6 - H_4 &= (2\epsilon_2 + \epsilon_3 + \epsilon_1) t_3
\end{aligned}
\label{diff-Ham-SU(3)}
\ee
 and this data is uniquely fixed by the compact part of the geometry.
 The relevant toric data can be encoded in the following pictures
\begin{figure}[ht]
\begin{center}
\includegraphics[scale=0.5]{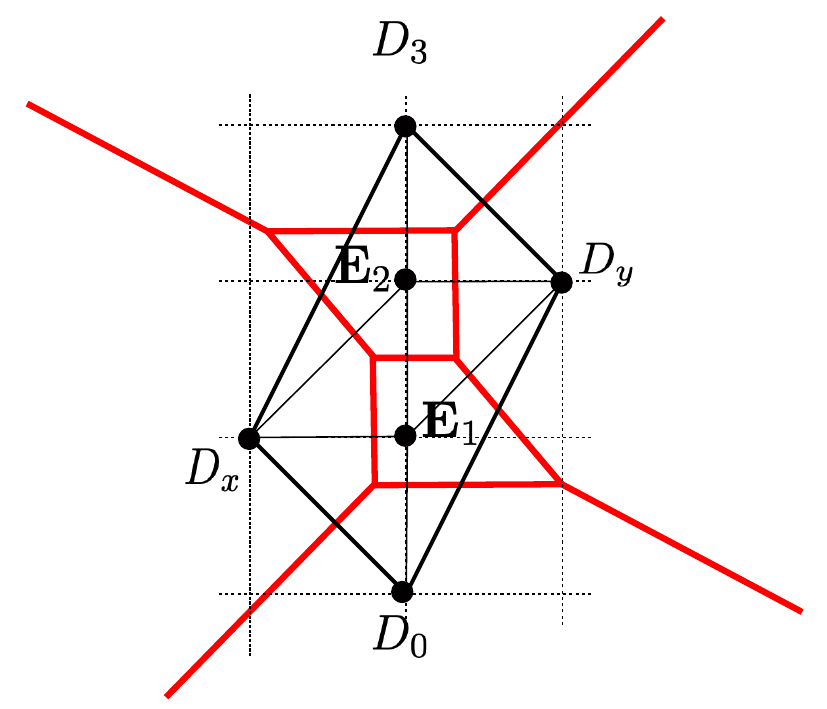}
\end{center}
\caption{The $SU(3)_0$ geometry.}
\label{fig:SU3o}
\end{figure}
\be
\begin{tikzpicture}
  \matrix (m) [matrix of math nodes,row sep=3em,column sep=8em,minimum width=2em]
  {
     v_1 & v_2 \\
     v_3 & v_4 \\
     v_5 & v_6 \\};
  \path[-]
    (m-1-1) edge node [left] {$t_2$} (m-2-1)
    	    edge node [above] {$t_1$} (m-1-2)
	    edge [draw=none] node [draw, shape=circle] {$1$} (m-2-2)
    (m-1-2) edge node [right] {$t_2$} (m-2-2)
    (m-2-1) edge node [above] {$t_1-t_2$} (m-2-2)
    	    edge node [left] {$t_3$} (m-3-1)
 	    edge [draw=none] node [draw, shape=circle] {$2$} (m-3-2)
    (m-2-2) edge node [right] {$t_3$} (m-3-2)
    (m-3-1) edge node [below] {$t_3+t_1-t_2$} (m-3-2);
\end{tikzpicture}
\qquad
\begin{tikzpicture}
  \matrix (m) [matrix of math nodes,row sep=3em,column sep=8em,minimum width=2em]
  {
     v_1 & v_2 \\
     v_3 & v_4 \\
     v_5 & v_6 \\};
  \path[-]
	  (m-1-1) edge node [left] {$-2$} node [right] {$0$} (m-2-1)
	  edge node [above] {$-3$} node [below] {$1$} (m-1-2)
	  (m-1-2) edge node [right] {$-2$} node [left] {$0$} (m-2-2)
	  (m-2-1) edge node [above] {$-1$} node [below] {$-1$} (m-2-2)
	  edge node [left] {$-2$} node [right] {$0$} (m-3-1)
	  (m-2-2) edge node [right] {$-2$} node [left] {$0$} (m-3-2)
	  (m-3-1) edge node [above] {$1$} node [below] {$-3$} (m-3-2);
\end{tikzpicture}
\ee
 where the first diagram labels vertices, edges and faces together with sizes of the edges in terms moment map data,
 while the second  diagram keeps track of $\psi$ data. Here we assume that  $t_1-t_2>0$.
The geometry has two compact faces (labeled by circles), so $m \in \BZ^{2n}$, and \cref{edge-face} becomes
\be
\begin{tabu}{c||c|c|c|c|c}
                  & Q_{12}       & Q_{34}                 & Q_{56}       & Q_{13}=Q_{24}           & Q_{35} = Q_{46} \\
\hline
t_e               & t_1          & t_1-t_2                & t_3+t_1-t_2  & t_2                     & t_3             \\
\psi \cdot m_\ell & -3m_{1,\ell} & -m_{1,\ell}-m_{2,\ell} & -3m_{2,\ell} & -2m_{1,\ell}+m_{2,\ell} & -2m_{2,\ell}+m_{1,\ell}
\end{tabu}
\ee
We have the following shifts of $\alpha$'s at each vertex
\be
\begin{aligned}
\alpha^{(1)} &= \tilde{\alpha} + m_1 \epsilon_1 \\
\alpha^{(2)} &= \tilde{\alpha} + m_1 (3 \epsilon_3 + \epsilon_1) \\
\alpha^{(3)} &= \tilde{\alpha} + m_1 (\epsilon_1 + 2 \epsilon_2) - m_2 \epsilon_2 \\
\alpha^{(4)} &= \tilde{\alpha} + m_1 ( 2\epsilon_2 + \epsilon_3 + \epsilon_1) + m_2 (\epsilon_3 - \epsilon_2) \\
\alpha^{(5)} &= \tilde{\alpha} + m_2 (2\epsilon_1 + 3 \epsilon_2) \\
\alpha^{(6)} &= \tilde{\alpha} + m_2 ( 2\epsilon_1 + 3 \epsilon_3 + 3 \epsilon_2)
\end{aligned}
\ee
where we suppressed the Lie algebra index for $U(n)$ gauge theory. 
The contribution of fluxes to classical terms can be computed using only \cref{table-gle-SU(3),diff-Ham-SU(3)}: 
\begin{multline}
\label{su3cf}
\begin{multlined}
	  - \sum_i   \sum_{v \in \Delta_X^{(0)}}  \frac{(\alpha_i^{(v)})^3}{3! \epsilon_1^{(v)} \epsilon_2^{(v)}\epsilon_3^{(v)}}
	  =  - \sum_i   \sum_{v \in \Delta_X^{(0)}}  \frac{\tilde\alpha_i^3}{3! \epsilon_1^{(v)} \epsilon_2^{(v)}\epsilon_3^{(v)}}\\
	  -   \sum_i \left ( \frac{4}{3} m_{1,i}^3 + \frac{4}{3}m_{2,i}^3 -
 \frac{1}{2} m_{1,i}^2 m_{2,i} -\frac{1}{2}  m_{1,i} m_{2,i}^2 \right )~,
\end{multlined}
\\
\begin{multlined}
-\sum_i \sum_{v\in\Delta_X^{(0)}} \frac{\alpha_i^{(v)}H^2_v}{2 \epsilon_1^{(v)} \epsilon_2^{(v)}\epsilon_3^{(v)}}
=-\sum_i \sum_{v\in\Delta_X^{(0)}} \frac{\tilde\alpha_i  H^2_v}{2 \epsilon_1^{(v)} \epsilon_2^{(v)}\epsilon_3^{(v)}} \\
-\left (t_1 t_2- \frac{1}{2}t_2^2  \right )  \sum_i m_{1,i}
   -\left  ((t_1- t_2) t_3 + \frac{1}{2}t_3^2  \right )  \sum_i m_{2,i}~,
\end{multlined}
\\
\begin{multlined}
- \sum_i  \sum_{v\in\Delta_X^{(0)}} \frac{(\alpha_i^{(v)})^2 H_v}{2 \epsilon_1^{(v)} \epsilon_2^{(v)}\epsilon_3^{(v)}}
= - \sum_i  \sum_{v\in\Delta_X^{(0)}} \frac{\tilde\alpha_i^2 H_v}{2 \epsilon_1^{(v)} \epsilon_2^{(v)}\epsilon_3^{(v)}}  \\
\quad - \frac{1}{2} \left ( 2t_1 + t_2 \right ) \sum_i m_{1,i}^2 -\left ( t_2 - t_1 \right ) \sum_i m_{1,i} m_{2,i}  - \frac{1}{2} \left ( 3 t_3 + 2 t_1 - 2 t_2 \right ) 
 \sum_i m_{2,i}^2~.
\end{multlined}
\end{multline}

    If we set $\varepsilon_4=\varepsilon_5=0$ in \cref{calFr} and perform the explicit computation 
\be
{\mathcal F} (t_1,t_2,t_3, \varepsilon) = -\frac{1}{3} t_1(t_2^2+t_2 t_3 + t_3^2) +\frac{1}{6} t_2(2t^2_2 + 2t_2 t_3 + 3t_3^2) +
f(\varepsilon) (t_1-2t_2 -t_3)^3~,\label{SU3-action-mod}
\ee
 where  
 \be
 f(\varepsilon)=
 \frac
 {\varepsilon_2^2 \varepsilon_3 \varepsilon_6 + \varepsilon_1^2 (\varepsilon_2 + \varepsilon_3) (\varepsilon_2 + \varepsilon_6) 
 + \varepsilon_1 \varepsilon_2 (\varepsilon_3 \varepsilon_6 + \varepsilon_2 (\varepsilon_3 + \varepsilon_6))}
   {18 (\varepsilon_6 + \varepsilon_1) (\varepsilon_6 + \varepsilon_2) (\varepsilon_1 + \varepsilon_3) (\varepsilon_2 + \varepsilon_3) }
 \ee
 As expected the function   ${\mathcal F} (t_1,t_2,t_3, \varepsilon)$ satisfies 
\begin{multline}
{\mathcal F} (t_1 +3m_1, t_2 - m_2 + 2m_1, t_3 - m_1 + 2m_2, \varepsilon) = {\mathcal F} (t_1, t_2, t_3, \varepsilon)  \\
+\left (-\frac{4}{3} m_1^3 - \frac{4}{3} m_2^3 +  \frac{1}{2} m_1^2 m_2 + \frac{1}{2} m_1 m_2^2 \right ) \\
+ \left ( m_1^2 (t_1 + \frac{1}{2} t_2) + m_2^2 (t_1 - t_2 + \frac{3}{2} t_3) + m_1 m_2 (t_2- t_1) \right ) \\
+\left ( m_1 (-t_1 t_2 + \frac{1}{2} t_2^2) + m_2 (t_2 t_3  - t_1 t_3 - \frac{1}{2} t_3^2) \right )
\label{shift-sym-SU(3)}
\end{multline}
to be compared with \cref{su3cf}. We focus on the universal part
\be
\mathcal F (t_1, t_2, t_3) =
-\frac{t_1 t_2^2}{3} + \frac{t_2^3}{3} - \frac{t_1 t_2 t_3}{3} + \frac{t_2^2 t_3}{2} - \frac{t_1 t_3^2}{3} + \frac{t_2 t_3^2}{2}~,\label{SU3-class-geom}
\ee
 although in what follows we only use the shift symmetry \cref{shift-sym-SU(3)}. 
 Finally we calculate   $\mathcal P$ in an analogous way to the $SU(2)$ case.
  For the given CY $\mathcal P$ is defined as
\begin{multline}
q_{123}^* \mathcal P_{(m_1,m_2)} (q_1, q_2, q_3)
= \frac{q^{m_1}_1 -1}{(1-q_1) (1-q_2) (1-q_3)}  + \frac{q_3^{3m_1} q_1^{m_1}-1}{(1-q_3^3q_1) (1-q_2 q_3^{-1}) (1-q_3^{-1})} \\
+ \frac{q_1^{m_1} q_2^{2m_1} q_2^{-m_2} -1}{(1-q_2^{-1})(1-q_1 q_2^2)(1-q_3)}
 + \frac{q_2^{2m_1} q_3^{m_1} q_1^{m_1} q_3^{m_2} q_2^{-m_2} -1}{(1-q_3 q_2^{-1})(1-q_2^2 q_3 q_1)(1-q_3^{-1})} \\
+ \frac{q_1^{2m_2} q_2^{3m_2} -1}{(1-q_1^{-1} q_2^{-2})(1-q_1^2 q_2^3)(1-q_3)} 
+ \frac{q_1^{2m_2} q_3^{3m_2} q_2^{3m_2}-1}{(1-q_2^{-2} q_3^{-1} q_1^{-1})(1-q_1^2 q_3^3 q_2^3)(1-q_3^{-1})}
\end{multline}
which has the property
 \be
 \mathcal P_{(-m_1,-m_2)} (q_1, q_2, q_3) =   -q_{123}^{-1} \mathcal P_{(m_1,m_2)} (q^{-1}_1, q^{-1}_2, q^{-1}_3)
 \ee
  Assuming $m_1 >0$ and $m_2 >0$ we can compute 
  \be
  \frac{ \mathcal P_{(m_1,m_2)} (q_1, q_2, q_3)}{q_{123}}
  = - \sum_{s=0}^{m_1-1} \sum_{k=0}^{2s} \sum_{l=0}^{3s-k} q_1^s q_2^k q_3^l  -
    \sum_{s=0}^{m_2-1} \sum_{k=0}^{2s} \sum_{l=0}^{s+k} q_1^k q_2^{2k-s} q_3^l +
    \sum_{s=0}^{m_1-1} \sum_{l=0}^{m_2-1} \sum_{k=0}^{s+l} q_1^s q_2^{2s-l} q_3^k
  \ee
and its size 
   \be
   |\mathcal P_{(m_1,m_2)}| =- \frac{1}{3} \left ( 4 m_1^3 - m_1 \right ) -  \frac{1}{3} \left ( 4 m_2^3 - m_2 \right ) + \frac{1}{2} m_1^2 m_2 +
    \frac{1}{2} m_1 m_2^2
   \ee
    for any integer $m_1$ and $m_2$.
    
Using the shift property \cref{shift-sym-SU(3)} and formulas from \cref{app:combin} we can write
\begin{multline}
\sum_{i=1}^n \frac {\mathcal F(t + \psi\cdot g_i m_i+\frac{\epsilon}2 \psi\cdot\sigma_i)} {g_i^2 - (\epsilon/2)^2}
= \frac{n \mathcal F(t + \psi \cdot g \frac{m_*}{n})} {g^2 - (n \epsilon/2)^2}
+\frac{\epsilon}2  \sum_{i<j} p(m_{1,ij}, m_{2,ij}) \\
+ g \sum_i \left(  p(m_{1,i}, m_{2,i}) -p(\frac{m_{1*}}{n}, \frac{m_{2*}}{n}) \right ) \\
 + \sum_i \left ( (m_{1,i}^2- \frac{m_{1*}^2}{n^2})  (t_1 + \frac{1}{2} t_2) + (m_{2,i}^2 - \frac{m_{2*}^2}{n^2})   (t_1 - t_2 + \frac{3}{2} t_3) \right.\\
 \left. + (m_{1,i} m_{2,i}- \frac{m_{1*}m_{2*}}{n^2}) (t_2- t_1) \right ) ~,\label{SU3-class-identity}
\end{multline}
where we defined
\be
p(m_1, m_2) = \frac43 m_1^3 + \frac43 m_2^3 - \frac12 m_1^2 m_2 - \frac12 m_1 m_2^2~. 
\ee
 The relation \cref{SU3-class-identity} can be written in other forms, e.g.\ there is a version 
  of formula \cref{SU2-alternative-rel} for this CY. 
 If we combine \cref{SU3-class-identity} with the properties of $|\mathcal P_{(m_1,m_2)}|$
\begin{multline}
u + \frac{\epsilon}2 \sum_{i<j} (s_{ij} + |\mathcal P_{(m_{1,ij}, m_{2,ij})}|)
=
\sum_i \frac{\mathcal F(t + g_i \psi \cdot m_i+\frac\epsilon2 \psi\cdot\sigma_i)}{g^2_i - (\epsilon/2)^2} \\
\quad
+\sum_i  \left[
g_i \left( \sum_v |K_{i,v}^{reg}|-\sum_e f_{\lambda_{i,e}} \right)  -\sum_{e} |\lambda_{i,e}| (t_e + g_i \psi \cdot m_i+\psi\cdot \sigma_i \frac{\epsilon}2)
	 \right] \\
\quad - \frac{1}{24} c_2(X)\cdot (nt + 2 \frac\epsilon2 \sum_{i<j} \psi \cdot m_{ij})~,
\end{multline}
where we used the property
\be
-\frac{1}{24} c_2(X) \cdot (t + \psi \cdot m) + \frac{1}{24} c_2(X) \cdot t
= \frac{m_1+m_2}6~,
\ee
which can be deduced from \cref{c2dt}.
 
 Finally we can use the geometrical engineering dictionary for $SU(3)$ theory by identifying 
  the K\"ahler parameters $(t_1,t_2,t_3)$ with two scalars $(\varphi_1, \varphi_2)$ in $SU(3)$ vector multiplet 
   and the coupling constant $h$ as
\be
t_1= h + 3 \varphi_1, \quad t_2 = 2 \varphi_1 - \varphi_2, \quad t_3 = 2 \varphi_2 - \varphi_1~.\label{SU3-dictionary}
\ee
We then match \emph{exactly} the 7d and 5d master formulas \cref{7d-master,5d-master} by identifying
\be
\begin{aligned}
 h_{1,\ell} &= -\sum_{i=1}^\ell m_{1,i} + \frac12 \sum_{i=1}^n  m_{1,i} = \frac12 \left( -m_{1,1}-\cdots-m_{1,\ell}+m_{1,\ell+1} + \cdots + m_{1,n} \right)~,\\
 h_{2,\ell} &= -\sum_{i=1}^\ell m_{2,i} + \frac12 \sum_{i=1}^n  m_{2,i} = \frac12 \left( -m_{2,1}-\cdots-m_{2,\ell}+m_{2,\ell+1} + \cdots + m_{2,n} \right)
\end{aligned}
\ee
and imposing the conditions $m_{1*} =m_{2*}= 0$, which imply  $h_{1,0}=h_{1,n}=0$, $h_{2,0}=h_{2,n}=0$ and amount to going from $U(n)$ to $SU(n)$ in 7d.
 We conclude that the partition function for 7d $SU(n)$ theory on the given CY is
  the same as the partition function for 5d $SU(3)_0$ theory on $A_{n-1}$ space (both theories are extended to $S^1$ in the appropriate fashion). 
 Using \cref{SU3-dictionary} the classical part \cref{SU3-class-geom} becomes
\be
\mathcal F
= - h(\varphi_1^2 + \varphi_2^2 - \varphi_1 \varphi_2 ) - \frac{4}{3} \varphi_1^3 - \frac{4}{3} \varphi_2^3 +
 \frac{1}{2} \varphi_1^2 \varphi_2 + \frac{1}{2} \varphi_1 \varphi_2^2
\ee
If instead we use \cref{SU3-action-mod} then we have 
   \be
    {\cal F} = - h(\varphi_1^2 + \varphi_2^2 - \varphi_1 \varphi_2 ) - \frac{4}{3} \varphi_1^3 - \frac{4}{3} \varphi_2^3 +
 \frac{1}{2} \varphi_1^2 \varphi_2 + \frac{1}{2} \varphi_1 \varphi_2^2  + f(\varepsilon) h^3~,
\ee
 which may correspond to adding some non-dynamical (purely geometric) term on 5d side.

\section{Conclusions and speculations}

Our main achievement in this paper is the 7d master formula,
which we derive resting on two claims.
The first, independence on 7d Coulomb branch parameters, has a deep meaning,
both mathematically (compactness) and physically (properties of an index of M-theory).
The second is more technical in nature, and has to do with factorization properties of $\mathcal F$.
A better (equivariant) understanding of $\mathcal F$ and its (shift) properties,
which will be discussed elsewhere, allows one to prove it.
These properties are due to the interplay of $\mathcal F$ with D4-branes
wrapping compact cycles, which play a crucial role in our correspondence.
For geometries that admit a geometric engineering,
the 7d master formula nicely matches the 5d one,
extending the geometric engineering paradigm from $A_0$ to $A_n$ geometries.

\subsection{\texorpdfstring{$TN_n$}{TNn} 5d instanton partition function}

Let us finish with a few remarks about the instanton partition function on $TN_n$ space.
We showed that $SU(n)$ 7d theory on CY is equivalent to 5d theory
(which is prescribed by a given CY)
on $A_{n-1}$ space with the following identification
\be
 m_{1,\alpha} = h_{0,\alpha}-h_{1,\alpha}, \ldots ,
 m_{n-1,\alpha} = h_{(n-2),\alpha} - h_{(n-1),\alpha},
 \text{ and } m_{n,\alpha} = h_{(n-1),\alpha} + h_{0,\alpha}
\ee
with
\be
 \sum_{i=1}^n m_{i,\alpha} = m_{*,\alpha} = 2 h_{0, \alpha}~,
\ee
where the parameter $\alpha$ stands for Cartan for 5d theory and $i=1,\ldots,n$.
For the case of 7d $SU(n)$ (5d on $A_{n-1}$)
we assume $h_{0,\alpha}=0$ and in this case
both $m_{i,\alpha}$ and $h_{i,\alpha}$ are integers.
For $U(n)$ 7d theory we drop the traceless condition for $m$'s
and the resulting theory should correspond to 5d theory on $TN_n$.
  If we take 7d master formula \cref{7d-master} and combine it with the above dictionary, we get 
   the following conjecture for 5d partition function on $TN_n$
\be
 Z_{SU(N)}^{5d} (TN_n \times S^1; z, \vec{b}, q_4, q_5) =
 \sum_{\vec{h}_0} e^{f(2\vec{h_0})} \sum_{(\vec{h}_1, \ldots, \vec{h}_{n-1})}
 \prod_{i=1}^n Z^{5d}_{SU(N)} ({\BC}^2 \times S^1; z, \vec{b}^{(i)} , q^{(i)}_4, q^{(i)}_5)~,
 \label{conj-5d-TN}
\ee
where the function $f$ is the same function that appears in \cref{7d-master},
$(q^{(i)}_4, q^{(i)}_5)$ are defined in \cref{5ddef-q4-q5}
and $\vec{b}^{(i)}$ in \cref{def-b5d}.
In the case $m_*\neq 0$ we cannot claim that $h_{i,\alpha}$ are integers
(but their appropriate differences are integers).
The function $f(m_*)=f(2h_0)$ is a cubic polynomial in $m_*$ ($h_0$)
and it can be calculated explicitly.
However, the concrete form of $f$ depends on 7d classical action \cref{our-def-classical},
e.g.~adding the term $g^{-1} ch_1$ to \cref{our-def-classical} simplifies $f$ a bit.
At the present level of understanding, for a given 7d classical action
we can calculate the polynomial $f$ explicitly.
However we do not understand what the 5d interpretation of this term is.
It is natural to expect that $f$ can be absorbed into classical 5d terms.
To illustrate this, let us rewrite $A_{n-1}$ case in \cref{5d-classical-An} for $TN_n$
\begin{multline}
 \beta^{-1} \log \Big (Z^{5d}_{\rm cl} (TN_{n}\times S^1)\Big ) =
 \sum_{i=1}^n \frac{\Big \langle \vec\varphi + \vec{h}_i \epsilon_4^{(i)}
  + \vec{h}_{i-1} \epsilon_5^{(i)}, \vec\varphi + \vec{h}_i \epsilon_4^{(i)}
  + \vec{h}_{i-1} \epsilon_5^{(i)} \Big \rangle}
 { \epsilon_4^{(i)} \epsilon_5^{(i)} } \\
 = \frac{\langle \vec\varphi + (\epsilon_5-\epsilon_4) \vec{h}_0, \vec\varphi
 + (\epsilon_5-\epsilon_4) \vec{h}_0\rangle }
 {n \epsilon_4 \epsilon_5} - \sum_{i=1}^n
 \Big \langle \vec{m}_i  - \frac{\vec{m}_*}{n}, \vec{m}_i - \frac{\vec{m}_*}{n} \Big \rangle~,
\label{last-5d-classical-TN}
\end{multline}
where $\langle~,~\rangle$ stands for the Lie algebra pairing.
This simple calculation is suggestive but at the moment we cannot claim
that we can do the same for all terms in $f$.
We expect the answer to take the form \cref{conj-5d-TN},
but we need a better 5d insight to fix ambiguities associated to $f$.

\subsection{Further directions}

It would be desirable to construct the full equivariant background in M-theory.
This would allow to completely fix the form of 7d classical action
and fully justify our constructions.
This background contains $G_4$ flux,
which technically implies certain shift symmetry properties for $\mathcal F$.
The fully equivariant definition of $\mathcal F$
(and of the twisted M-theory $\Omega$-background)
and its interplay with $H^2$ vs $H^2_c$ is something we plan to address in the future.

We could replace $\BC^2/\BZ_n$ with a more general $\Gamma_\mathfrak{g} \subseteq SU(2)$,
which by the McKay correspondence is classified by ADE
\be
\begin{tabu}{c|ccccc}
\Gamma_\mathfrak{g} & \BZ_n & \mathbb D_n & \mathbb T & \mathbb O & \mathbb I \\
\hline
\mathfrak{g} & \mathfrak{su}(n) & \mathfrak{so}(2n) & \mathfrak{e}_{6} & \mathfrak{e}_{7} & \mathfrak{e}_{8}
\end{tabu}
\ee
although the DT counterpart of this has not been fully developed.\footnote
{In particular, it is unclear whether these will produce genuinely new invariants or not,
which makes the question worth investigating.}
Here $\BZ_n$ is the $n$-th cyclic group,
$\mathbb D_n$ is the $n$-th binary dihedral group,
$\mathbb T$ is the binary tetrahedral group,
$\mathbb O$ is the binary octahedral group,
and $\mathbb I$ is the binary icosahedral group.

More intriguing examples of our relations occur if we consider a hybrid setup
for which one of the two manifolds is compact and the other is non-compact.
For instance consider the case $M_4 = S^4$.
On one side we have the index of a 5d SCFT,
on the other we have the index of the 7d gravitational theory on $S^1 \times M_6^\sharp$,
where $\sharp$ denotes resolution.
Perhaps even more interesting is the case $M_4 = K3$,
where we could learn about the physics of M-theory on K3 from studying partition functions of 5d SCFTs.

\subsection*{Acknowledgements}

The authors thank Sergey Cherkis, Cyril Closset, Lorenzo Foscolo, Guglielmo Lockhart,
Joseph Minahan and Francesco Sala for discussions.
The work of MDZ has received funding from the European Research Council (ERC)
under the European Union’s Horizon 2020 research and innovation programme (grant agreement No. 851931).
The work of NP and MZ is supported by the grant ``Geometry and Physics"
from the Knut and Alice Wallenberg foundation.
Opinions and conclusions expressed here are those of the authors
and do not necessarily reflect the views of funding agencies.

\appendix

\section{ALE spaces of \texorpdfstring{$A_{n-1}$}{An-1} type}\label{App:An-1}
  
  We collect basic information about ALE spaces of type $A_{n-1}$. They are
   hyperK\"ahler manifolds that are the deformation (resolution) of $\BC^2/\BZ_n$.
   We are interested in their toric geometry. 
  
    \subsection{\texorpdfstring{$A_1$}{A1} space}
  
  We start with the simplest example, namely $A_1$ type.
   First consider the singular space  $\BC^2/{\BZ}_2$ with  $\BZ_2$-action on $\BC^2$ 
   $(z_1, z_2) \rightarrow (-z_1, -z_2)$. We can define invariant coordinates
   \be
    x =  z_1^2 ~,~~~~~y = z_2^2~,~~~~~w=z_1 z_2
   \ee
    with the relation
    \be
     xy - w^2=0~,
    \ee
     which defines the singular $A_1$ space as a condition in $\BC^3$. 
    Alternatively we can define this space as quotient of $\BC^3= (z_1, z_2, z_3)$ by $\BC_*$-action with charges $(1, -2, 1)$
     \be
     (z_1, z_2, z_3 )~~\rightarrow~~(  \lambda z_1, \lambda^{-2} z_2, \lambda z_3)
    \ee
     and introduce invariant coordinates 
     \be
      x= z_1^2 z_2~,~~~~~
      y=z_3^2 z_2~,~~~~~
      w= z_1 z_2 z_3
     \ee
      subject to the same condition in $\BC^3$
      \be
      xy - w^2 =0~.
      \ee
       The way to resolve this space is to remove the point $z_1=z_3=0$ and thus the resulting space is $O(-2) \rightarrow \mathbb{CP}^1$, which is the same as $T^*\mathbb{CP}^1$.
        This space is equipped with the well-known Eguchi-Hanson metric (hyperK\"ahler metric) and it can be obtained either as K\"ahler reduction of $\BC^3$ or as hyperK\"ahler 
         reduction of $\BC^4$. 
      
        Since we are interested in the  toric geometry of this space let us concentrate on the K\"ahler quotient picture.  We can obtain this space by the K\"ahler quotient 
       of $\BC^3$ with respect to $U(1)$ acting with charges $(1,-2,1)$. The corresponding moment map is 
       \be
        |z_1|^2  -2 |z_2|^2  +  |z_3|^2 = t 
       \ee
        with $t>0$, which is related to the size of $\mathbb{CP}^1$. The case $t=0$ corresponds to the singular space.  The resulting space 
         $A_1$ can be covered by two patches with coordinates 
\be 
\begin{aligned}
 1 & \quad \Big (\xi_1^{(1)} = z_1^2 z_2 ,~  \xi_2^{(1)}=\frac{z_3}{z_1} \Big )~, \\
 2 & \quad \Big ( \xi_1^{(2)} =\frac{z_1}{z_3}, ~\xi_2^{(2)}= z_3^2 z_2 \Big )~,
\end{aligned}
\ee
where on patch 1 we assume $z_1 \neq 0$ and on patch 2 $z_2 \neq 0$. On the intersection of two patches we have 
          the coordinate change
          \be
	  \xi_1^{(2)} = \frac{1}{\xi_2^{(1)}}~, \quad \xi_2^{(2)} = (\xi_1^{(2)})^{-2} \xi_1^{(1)}~,
	   \label{A1-transform}
          \ee
          which confirms that we deal with $O(-2)$ bundle over $\mathbb{CP}^1$. 
      There is a $T^2$-action on $A_1$ with two fixed points: patch 1 contains $(\xi_1^{(1)}, \xi_2^{(1)})=(0,0)$ (in $\BC^3$-coordinates $z_3=0, z_2=0$) and patch 2 contains $(\xi_1^{(2)}, \xi_2^{(2)})=(0,0)$ (in $\BC^3$-coordinates  $z_1=0, z_2 =0$).
If we define a $\mathbb{T}^3$-action on $\BC^3$ as $z_i \rightarrow e^{\iu \varepsilon_i} z_i$, we can read off the $T^2$ action on the homogeneous coordinates
\be
\begin{aligned}
1& \quad \xi_1^{(1)} \rightarrow e^{i(2\varepsilon_1 + \varepsilon_2)}  \xi_1^{(1)}  ~,\quad \xi_2^{(1)}  \rightarrow e^{i(\varepsilon_3 - \varepsilon_1)}  \xi_2^{(2)} ~, \\
2& \quad \xi_1^{(2)} \rightarrow e^{i(\varepsilon_1 - \varepsilon_3)}  \xi_1^{(2)}~,\quad \xi_2^{(2)} \rightarrow e^{i(2\varepsilon_3 + \varepsilon_2)} \xi_2^{(2)}~,
\end{aligned}
\ee
and since we deal only with $T^2$-action it is convenient to define two independent $(\epsilon_4, \epsilon_5)$ such that 
 $2 \epsilon_4= 2\varepsilon_1 +\varepsilon_2$ and $2\epsilon_5 = 2 \varepsilon_3 + \varepsilon_2$ or alternatively we can set $\varepsilon_2=0$ with 
  the identification $\epsilon_4= \varepsilon_1$ and $\epsilon_5 = \varepsilon_3$.
      Therefore the fixed point data for the two fixed points is given by 
      \be
       (2\epsilon_4, \epsilon_5 - \epsilon_4)~,\quad(\epsilon_4 - \epsilon_5, 2\epsilon_5)~. 
      \ee
     If on ambient space $\BC^3$ we define the standard Hamiltonian $H = \varepsilon_1 |z_1|^2 +\varepsilon_2 |z_2|^2 + \varepsilon_3|z_3|^2$
      then we can evaluate its values at the fixed points of the quotient space. Using the DH theorem we can evaluate the  equivariant volume of the quotient space
       as follows
       \be
        {\rm vol} (A_1) = \frac{e^{H_1}}{(\epsilon_5 - \epsilon_4) 2 \epsilon_4} + \frac{e^{H_2}}{(\epsilon_4 - \epsilon_5) 2 \epsilon_5} = \frac{1}{2\epsilon_4 \epsilon_5} - \frac{1}{4} t^2  - \frac{1}{12} (\epsilon_4 + \epsilon_5) t^3 + O(\epsilon^2)~,
       \ee
       where $H_1=\epsilon_4 t$ and $H_2=\epsilon_5 t$ are the values of Hamiltonian at the fixed points.

    \subsection{\texorpdfstring{$A_{n-1}$}{An-1} space} 
   
  We consider ALE spaces of type $A_{n-1}$.  The singular $A_{n-1}$ space corresponds to 
    $\BC^2/\BZ_{n}$ with $\BZ_{n}$ generated by ${\rm diag} (e^{\frac{2\pi  \iu}{n}}, e^{-\frac{2 \pi \iu}{n}})$ acting on $(z_1, z_2)$. 
   We can define invariant coordinates 
   \be
  x =  z_1^{n} ~,~~~~~y = z_2^{n}~,~~~~~w=z_1 z_2
   \ee
   and realize $A_{n-1}$ singular space in $\BC^3$ as
    \be
     xy - w^{n}=0~. 
    \ee
      Alternatively, we can think of $A_{n-1}$ as $\BC^{n-1}_*$-quotient of $\BC^{n+1}$ with charges
     \be
Q=
\begin{pmatrix}
1 & -2 & 1 & 0 & \cdots & 0 & 0 & 0 & 0 \\
0 & 1 & -2 & 1 & \cdots & 0 & 0 & 0 & 0 \\
  \cdots &  \cdots  &  \cdots  &  \cdots  & \cdots &  \cdots  &  \cdots  &  \cdots  &  \cdots \\
   0 & 0 & 0 & 0 & \cdots & 1 & -2 & 1 & 0 \\
    0 & 0 & 0 & 0 & \cdots & 0 & 1 & -2 & 1
\end{pmatrix}~.
\label{An-1-charges}
\ee      
Except for the first and last columns, this is (minus) the Cartan matrix of $A_{n-1}$.
Assuming for $\BC^{n+1}$ coordinates 
  $(z_1, z_2, \cdots z_{n+1})$, we introduce invariant coordinates 
\be
\begin{aligned}
    x &= z_1^{n} z_2^{n-1} z_3^{n-2} \cdots z_{n+1}^0~, \\
  y &= z_1^0 z_2^1 z_3^2 \cdots z_{n+1}^{n}~,\\
  w &= z_1 z_2 z_3 \cdots z_{n+1}~, 
\end{aligned}
\ee
 which satisfy the relation in $\BC^3$
        \be xy-w^{n}=0~. \ee
   We are interested in the toric geometry of the resolved $A_{n-1}$ space, which can be realized as K\"ahler quotient 
    $\BC^{n+1}//U(1)^{n-1}$ with charge matrix \cref{An-1-charges} and moment maps
       \be
   |z_\alpha|^2 - 2 |z_{\alpha+1}|^2 + |z_{\alpha +2}|^2 = t_{\alpha}~,~~~~\alpha =1,2, \cdots , n-1~,\label{moment-values}
  \ee
   with $t_\alpha >0$ for the resolved $A_{n-1}$ space. The resolved $A_{n-1}$ space can be covered by $n$ patches 
    with the following homogeneous coordinates
   \be
    \Big ( \xi_1^{(i)} = \frac{z_1^n z_2^{n-1} \cdots z_{n+1}^0}{(z_1 z_2 \cdots z_{n+1})^{i-1} }~,~~
    \xi_2^{(i)} = \frac{z_1^0 z_2^1 \cdots z_{n+1}^n}{(z_1 z_2 \cdots z_{n+1})^{n-i}} \Big ) ~,~~~~~i=1,2, \cdots , n~,
   \ee
     where all $z$'s are assumed to be non-zero except $z_{n+2-i}$ and $z_{n+1-i}$.
     At the intersection of patches  $i$  and $(i+1)$ we have the following coordinate change
        \be
     \xi_1^{(i+1)} = \frac{1}{\xi_2^{(i)}}~,\quad \xi_2^{(i+1)} = (\xi_2^{(i)})^2 \xi_1^{(i)} ~,
     \label{local-CP1}
    \ee
    which should be compared to the case of $O(-2)$-bundle over $\mathbb{CP}^1$, see \cref{A1-transform}. 
     The space $A_{n-1}$ admits a $T^2$-action with $n$ fixed points, with each patch $i$ containing fixed 
      point $(\xi_1^{(i)}, \xi_2^{(i)}) = (0,0)$ (or in $\BC^{n+1}$ coordinates $z_{n+2-i}=0$ and $z_{n+1-i}=0$).
    Between fixed points on patch $i$ and the nearby patch $(i+1)$ there is a $\mathbb{CP}^1$ as can be seen from the coordinate 
     transformations \cref{local-CP1}.  Let us work out how $T^2$ acts around every fixed point by analyzing the quotient.
  Assuming $T^{n+1}$ action on $\BC^{n+1}$ as $z_j \rightarrow e^{\iu \varepsilon_j} z_j$ with $j=1,2,\cdots, n+1$
   we can derive the toric action on the coordinates on patch $i$
\be
\begin{aligned}
 \xi_1^{(i)} & \rightarrow e^{\iu [ (n-i+1) \varepsilon_1 + (n-i) \varepsilon_2 + \cdots + (1-i) \varepsilon_{n+1}]} ~\xi_1^{(i)}~,\\
 \xi_2^{(i)} & \rightarrow e^{\iu [(i-n) \varepsilon_1 + (i-n+1) \varepsilon_2 + \cdots + i \varepsilon_{n+1}]} ~\xi_2^{(i)}~.
\end{aligned}
\ee
Since the resulting symmetry is just $T^2$ we can choose two independent parameters for the global $T^2$. 
   One can do the following choice $\varepsilon_2=...=\varepsilon_n =0$ with the identification\footnote
   {At the present level of discussion this choice looks ad hoc. However, there exists a proper treatment without arbitrary choices that essentially gives the same result. We will present it elsewhere.}
  \be
   \epsilon_4 =  \varepsilon_1~,~~~~~\epsilon_5=   \varepsilon_{n+1}~.
 \ee
   Using these global $\epsilon_4$ and $\epsilon_5$ we can derive  the following local action
 \be
  \epsilon_4^{(i)} = (n-i + 1) \epsilon_4 + (1-i) \epsilon_5~,~~~~~\epsilon_5^{(i)} = (i-n) \epsilon_4 + i \epsilon_5~,
  \label{An-local-global}
 \ee
  where we use notation $\xi_1^{(i)} \rightarrow e^{\iu \epsilon_4^{(i)}} \xi_1^{(i)}$ and $\xi_2^{(i)} \rightarrow e^{\iu \epsilon_5^{(i)}} \xi_2^{(i)}$ 
   around fixed point $(\xi_1^{(i)}, \xi_2^{(i)}) = (0,0)$. 
 
We are interested in the calculation of the equivariant volume using DH theorem
 \be
        {\rm vol} (A_{n-1}) = \sum_{i=1}^n \frac{e^{H_i}}{\epsilon_4^{(i)} \epsilon_5^{(i)} }  ~,\label{defin-equiv-An}
 \ee
    where $H_i$ is the value of Hamiltonian at the fixed point $i$ ($i=1,2,...,n$). 
    If on ambient space $\BC^{n+1}$ we define the Hamiltonian $H=\sum_{i=1}^{n+1}
    \varepsilon_i |z_i|^2$ then using above choices and the description of the fixed 
     points we can derive the value of Hamiltonian at the fixed point $i$ in 
      terms of the values of the moment maps \cref{moment-values}
     \be
  H_i= \epsilon_4 \sum_{j=1}^{n-i} j t_j + \epsilon_5 \sum_{j=1}^i (j-1) t_{n-j+1}~, \label{value-Hamil}
 \ee
 where we have $(n-1)$ $t$'s.
 Introduce $\alpha_i$ with $i=1,2,...,n$ such that
\be
 t_{n-i}=\alpha_{i+1} - \alpha_i  ~, \label{t-alpha-eqs}
\ee
 This map is not invertible unless we add an extra condition.
 It is natural to require 
 \be
 \sum_{i=1}^n\alpha_i=0~.
 \label{trace-alpha}
 \ee
 One can check that \cref{t-alpha-eqs,trace-alpha} provide an invertible map between $t$'s and $\alpha$'s. 
Using \cref{defin-equiv-An} with the values of $H_i$ in \cref{value-Hamil} expressed in terms of $\alpha$'s, we get
\be
        {\rm vol} (A_{n-1}) = \frac{1}{n\epsilon_4 \epsilon_5} - \frac{1}{2} \sum_{i=1}^n\alpha_i^2 + \frac{\epsilon_4+\epsilon_5}{12} \sum_{i<j} (\alpha_i -
        \alpha_j)^3 + \frac{n(\epsilon_4-\epsilon_5)}{12} \sum_{i=1}^n \alpha_i^3 + O(\epsilon^2)~.
\label{volAn}
\ee 
 The first two terms on RHS were derived in \cite{Moore:1997dj}.
 The cubic term in $\alpha$'s has a nice form.

\section{\texorpdfstring{$A_{n-1}$}{An-1} vs \texorpdfstring{$TN_n$}{TNn}}
\label{App:TN}
 
  We collect information about the relation between the cyclic ALE spaces  and ALF spaces.  
 
 The ALE space of type $A_{n-1}$ is the four-dimensional hyper-K\"ahler manifold obtained by the hyper-K\"ahler reduction of $\mathbb{H}^n \times \mathbb{H}$   
  with respect to $U(1)^n$ acting as
  \be
 q_a \rightarrow q_a e^{i t_a}~,~~~~~~w \rightarrow w e^{i \sum_{a=1}^n t_a}~. 
\ee
 The resulting metric is of the form
\be
 ds^2_{A_{n-1}}   = \frac14 \tilde{V} d\boldsymbol r^2 + \frac14 \tilde{V}^{-1} (d\tau +\chi)^2
\ee
 where $\boldsymbol{r} \in \BR^3$,  $\tau$ is periodic with period $4\pi$, and $\boldsymbol x_a$ are the center's positions in $\BR^3$, such that $\boldsymbol x_a \neq \boldsymbol x_b$ ($a\neq b$) for non-singular space.
We also use the following notations
\be
 \tilde{V} = \sum_{a=1}^{n} V_a~,
 \quad V_a = \frac{1}{| \boldsymbol x_a - \boldsymbol r |}
\ee
and
\be
 \chi = \sum_{a=1}^n \chi_a~,
 \quad d\chi_a = \star_3 dV_a~.
\ee
For the case $n=1$ we recover the usual flat metric on $\BC^2$
and thus we denote $A_0 = \BC^2$.
For the case $n=2$ the above metric is the well-known Eguchi-Hanson metric on $T^* \mathbb{CP}^1$.

The cyclic ALF space, better known as multi-Taub-NUT space $TN_n$,
is the four-dimensional hyper-K\"ahler manifold obtained by hyper-K\"ahler reduction
of $\mathbb{H}^n \times \mathbb{H}$ wrt $\BR^n$ acting as
\be
 q_a \rightarrow q_a e^{i t_a}~,\quad w \rightarrow w + R \sum_{a=1}^n t_a~,
\ee
with $R > 0$.
The resulting metric has the form
\be
 ds^2_{TN_n} = \frac14 V d\boldsymbol r^2 + \frac14 {V}^{-1} (d\tau +\chi)^2
\ee
with
\be
  V = \frac{1}{R^2} + \tilde{V}~,
\ee
where we use the same notations as before.
Unlike $ds^2_{A_{n-1}}$ the metric on $TN_n$ is not asymptotically euclidean,
instead at infinity it approaches $\BR^3 \times S^1$
with $R$ being the radius of the circle.
In the limit $R \rightarrow \infty$ the metric $ds^2_{TN_n}$ goes to $ds^2_{A_{n-1}}$.

As hyper-K\"ahler manifolds $TN_n$ and $A_{n-1}$ are different,
however as holomorphic symplectic manifolds (i.e., a complex manifold with symplectic $(2,0)$ form)
they are the same \cite{MR1128554}.
Both metrics $ds^2_{TN_n}$ and $ds^2_{A_{n-1}}$ admit $T^2$-isometries
with one particular $U(1)$ being tri-holomorphic.
We are interested in $T^2$ action on $A_{n-1}$ and $TN_n$.
The detailed discussion of $T^2$ action on $A_{n-1}$ space has been presented in the previous appendix.
Despite the fact that $TN_n$ is not toric (i.e., it cannot be glued from affine $\BC^2$ patches)
we believe that our discussion of $T^2$ action around fixed points (in particular \cref{An-local-global}) goes through,
since our previous analysis involves only complex coordinates and
as complex manifolds these two spaces are the same.
Intuitively this is clear since close to the origin
(assuming that all centers $\boldsymbol x_a$ are close to the origin),
$TN_n$ is approximated by $A_{n-1}$.

Let us review some basic facts about the cohomologies of $TN_n$/$A_{n-1}$
and the line bundles over these spaces, following Witten \cite{Witten:2009xu}.
The space $TN_n$ has two types of interesting cycles:
compact 2-cycles $C_{a,b} \cong S^2$,
which are fibered over the line segments joining the points
with coordinates $\boldsymbol x_a$ and $\boldsymbol x_b$ in $\BR^3$,
and non-compact 2-cycles $C_a$ ($a=1,2,\ldots, n$).
On $TN_n$ there are two versions of homology.
The first version is topological $H_2 (TN_n, \BZ) = \BZ^{n-1}$,
which is dual to the compactly supported cohomology $H^2_{\rm cpct} (TN_n, \BZ)$.
Among all compact 2-cycles $C_{a,b}$ only $(n-1)$ are homologically independent
and we can pick the standard basis
\be
 D_a = C_{a, a+1}~, \quad a=1,2, \ldots, n-1
\ee
The intersection matrix is minus the Cartan matrix for $A_{n-1}$ group
     \be
     ({\cal C})_{ab} =      \begin{pmatrix}
 -2 & 1 & 0 & \cdots & 0 & 0 & 0 & 0 \\
  1 & -2 & 1 & \cdots & 0 & 0 & 0 & 0 \\
  \cdots &  \cdots  &  \cdots  &  \cdots  & \cdots &  \cdots  &  \cdots  &  \cdots  &  \\
   0 & 0 & 0 & 0 & \cdots & 1 & -2 & 1  \\
    0 & 0 & 0 & 0 & \cdots & 0 & 1 & -2 
\end{pmatrix}~.
\label{matrix-An-1}
\ee
The second version is ``geometrical'' homology $H_2 (TN_n, \BZ) = \BZ^n$,
which is generated by the non-compact cycles $C_a$
with intersection matrix $\langle C_a, C_b \rangle = \delta_{ab}$.
We can define the following curvature two form $B_a=d \Lambda_a$, with
\be
 \Lambda_a = \frac{1}{2} \chi_a - \frac{V_a}{2V} (d\tau + \chi)~,
\ee
The $B_a$ are of $(1,1)$-type (so anti-self dual) and
\be
 \frac{1}{2\pi} \int_{C_a} B_b = \delta_{ab}~.
\ee
Alternatively we have
\be
 \frac{1}{2\pi} \int_{C_{a,c}} B_b = \delta_{ab} - \delta_{bc}~, \quad
 \frac{1}{2\pi} \int_{D_a} (B_b - B_{b+1}) ={\cal C}_{ab} ~.
\ee
The curvature $B_a$ defines a line bundle ${\cal L}_a$ (correspondingly $m_a B_a$ defines ${\cal L}_a^{m_a}$). If we look at the sum 
   $B=\sum_{a=1}^n B_a$ then $B$ has vanishing integral over each compact cycle. However $B$ is a normalizable harmonic two form and thus  it is  non-trivial in $L^2$-cohomology \cite{Hausel:2002xg}.
    If we take the limit $R \rightarrow \infty$, then the form $B$ is not normalizable on $A_{n-1}$ 
     and there is no additional element in cohomology. Thus if we want to calculate the following integral
     \be
      \int_{TN_n} c_1^2 ({\cal L}) = \sum_{a=1}^n m_a^2
     \ee
     where ${\cal L} = \oplus_{a=1}^m {\cal L}_a^{m_a}$ then the main difference between $TN_n$ and $A_{n-1}$ is the trace condition $\sum_a m_a =0$. 
      On $A_{n-1}$ we have to impose the trace condition $\sum_a m_a =0$ since $B$ is not normalizable and so it is not an element of $L^2$-cohomology.

\section{Useful combinatorial identities}
\label{app:combin}
 
 In this appendix we collect the useful combinatorial identities that we use in the paper. 
 
 If we have two sequences of numbers $c_i$ and $d_i$ ($i=1,...,n$) the double sum 
  can be reduced to a single sum as follows
\be
\label{ccdd}
\sum_{i<j} (c_i+c_j)(d_i-d_j)
=
\sum_\ell c_\ell (d_1 +\ldots +d_{\ell-1} + d_\ell (n -2\ell +1)
- d_{\ell+1} - \ldots - d_n)~. 
\ee
 
For the sequence of $m_i$ we define the short hand notation $m_{ij}=m_i - m_j$ and we have 
\be
\sum_{i<j} m_{ij}
= \sum_{i=1}^n m_i (n+1-2i)
= \sum_{i=1}^n \sigma_i (m)~,
\ee
 where $\sigma_i (m)$ is defined in \cref{def-sigma}. 
If we define $g_i$ as in \cref{gi} we have
\be
\begin{aligned}
\sum_{i=1}^n
\frac{(g_i m_i+\frac{\epsilon}2 \sigma_i)^2} {g_i^2 - (\epsilon/2)^2}
	  &=
\frac{n \epsilon^2}{4g^2-(n\epsilon)^2} m_*^2
+\sum_{i=1}^n m_i^2 \\
\sum_{i=1}^n
\frac{g_i m_i+\frac{\epsilon}2 \sigma_i} {g_i^2 - (\epsilon/2)^2}
	  &=
\frac{g m_*} {g^2 - (n \epsilon/2)^2} \\
\sum_{i=1}^n
\frac{1} {g_i^2 - (\epsilon/2)^2}
	  &=
\frac{n} {g^2 - (n \epsilon/2)^2}
\end{aligned}
\ee  
  and
\be
\sum_{i=1}^n
\frac{(g_i m_i+\frac{\epsilon}2 \sigma_i)^3} {g_i^2 - (\epsilon/2)^2}
=
\frac{\epsilon}2 \sum_{i<j} m_{ij}^3
+\frac{\epsilon^2 g}{4g^2-(n\epsilon)^2} m_*^3
+g \sum_{i=1}^n m_i^3~,
\ee
 where $m_*=\sum_i m_i$. If we have two sequences $m_{i,a}$ and $m_{i,b}$ (in our context the labels $a,b$ are related to the faces of toric CY) then we have the following identities 
 \be
\sum_{i=1}^n
\frac{(g_i m_{i,a} +\frac{\epsilon}2 \sigma_i(m_a))  (g_i m_{i,b} +\frac{\epsilon}2 \sigma_i(m_b)) } {g_i^2 - (\epsilon/2)^2}  =
\frac{n \epsilon^2}{4g^2-(n\epsilon)^2} m_{*,a} m_{*,b}
+\sum_{i=1}^n m_{i,a} m_{i,b}
\ee
 and 
\begin{multline}
\sum_{i=1}^n
\frac{(g_i m_{i,a} +\frac{\epsilon}2 \sigma_i(m_a))^2  (g_i m_{i,b} +\frac{\epsilon}2 \sigma_i(m_b)) } {g_i^2 - (\epsilon/2)^2}\\
=
\frac{\epsilon}2 \sum_{i<j} m_{ij,a}^2 m_{ij,b}
+\frac{\epsilon^2 g}{4g^2-(n\epsilon)^2} m_{*,a}^2 m_{*,b}
+g \sum_{i=1}^n m_{i,a}^2 m_{i,b}
\end{multline}
with $m_{*,a}=\sum_i m_{i,a}$ and $m_{*,b}=\sum_i m_{i,b}$.

\printbibliography
\end{document}